\pdfoutput=1
\documentclass[useAMS,usenatbib,fleqn]{mn2e}

\usepackage{graphicx}
\usepackage{float}
\usepackage{epstopdf}
\usepackage{amsmath}
\usepackage{amssymb}
\usepackage[labelformat=empty,labelsep=none]{subcaption}
\usepackage{epsfig}
\usepackage{rotating}

\def\gtsim {>\kern-1.2em\lower1.1ex\hbox{$\sim$}~}   
\def\ltsim {<\kern-1.2em\lower1.1ex\hbox{$\sim$}~}   

\newcommand{\er}{\epsilon_\textrm{r}}
\newcommand{\ef}{\epsilon_\textrm{f}}
\newcommand{\ms}{M_\textrm{seed}}
\newcommand{\rc}{\rho_\textrm{c}}

\title[Seeding Black Holes in Cosmological Simulations]{Seeding Black Holes in Cosmological Simulations}
\author[P.~Taylor and C.~Kobayashi]{P.~Taylor\thanks{E-mail:
p.taylor7@herts.ac.uk} and C.~Kobayashi\\
Centre for Astrophysics Research, Science and Technology Research Institute, University of Hertfordshire, AL10 9AB, UK}

\begin{document}

\date{Accepted  Received ; in original form}

\pagerange{\pageref{firstpage}--\pageref{lastpage}} \pubyear{}

\maketitle

\label{firstpage}

\begin{abstract}
We present a new model for the formation of black holes in cosmological simulations, motivated by the first star formation.
Black holes form from high density peaks of primordial gas, and grow via both gas accretion and mergers.
Massive black holes heat the surrounding material, suppressing star formation at the centres of galaxies, and driving galactic winds.
We perform an investigation into the physical effects of the model parameters, and obtain a `best' set of these parameters by comparing the outcome of simulations to observations.
With this best set, we successfully reproduce the cosmic star formation rate history, black hole mass -- velocity dispersion relation, and the size -- velocity dispersion relation of galaxies.
The black hole seed mass is $\sim10^3M_\odot$, which is orders of magnitude smaller than has been used in previous cosmological simulations with active galactic nuclei, but suggests that the origin of the seed black holes is the death of Population {\sc iii} stars.

\end{abstract}



\section{Introduction}
\label{sec:intro}
While the evolution of dark matter in the standard $\Lambda$ cold dark matter (CDM) cosmology is reasonably well understood, the evolution of the baryonic component is much less certain because of the complexity of the relevant physical processes, such as star formation and feedback.
One approach to studying the dynamics of baryons in galaxy formation is to use semi-analytic techniques \citep[e.g.,][]{kauffmann93, cole94} that combine the growth of dark matter haloes with subgrid modelling of the baryonic physics \citep{fanidakis11,lagos13,fu13}.
With semi-analytic models, it has been shown that the observed luminosity function of galaxies can be reproduced with feedback both from supernovae and active galactic nuclei (AGN), where the numbers of low-mass and massive galaxies are decreased with supernova and AGN feedback, respectively \citep[e.g.,][]{croton06}.

The other approach is to directly implement star formation and feedback into hydrodynamical simulations.
Many simulation codes have been developed for studying galaxy formation and evolution, not only of isolated systems \citep[e.g.,][]{katz92,mihos94, steinmetz94, kawata03, ck04} but also for cosmological simulations of individual galaxies \citep[e.g.,][]{navarro94} or of the galaxy population as a whole \citep{cen99,springel03}.
With these simulations, it is possible to predict not only averaged properties of galaxies but also internal structures of galaxies, i.e., kinematics and spatial distribution of stars and gas within galaxies, which have also become available in recent observations with integral field units and multi-object spectrographs.
Given the computational resources, it is not possible to resolve star-forming clouds, supernova ejecta, and AGN, and therefore some parametrizations are inevitable.
However, most of the parameters can be constrained from observations, and we aim to do this in this paper.

Supernova and hypernova feedback play an important r\^ole in reducing the peak rate of cosmic star formation and reproducing mass-metallicity relations \citep[e.g.,][]{ck07}.
However, the `down-sizing' phenomenon is not fully reproduced \citep{cowie96,bundy06}.
Down-sizing has observationally been seen as the redshift evolution of the mass of galaxies with active star formation, the mass and redshift dependencies of the specific star formation rates \citep[SFRs;][]{juneau05,stark13}, the fundamental plane \citep{treu05}, and the [$\alpha$/Fe] ratios of early type-galaxies \citep{thomas05}.
In hydrodynamical simulations, because most of the stars in massive galaxies have formed in subgalaxies at high-redshift before they merge to the present galaxies, the stellar populations of massive galaxies are older than low-mass galaxies (\citealt{ck07}; see also \citealt{delucia06,fontanot09} for semi-analytic models). 
However, the present star formation rates are still too high in massive galaxies, and thus the specific star formation rates are also too high, and the [$\alpha$/Fe] ratios are too low for massive early-type galaxies.
Some additional feedback that can efficiently work in massive galaxies is required, and AGN feedback is the most plausible.

The importance of black holes has been underscored by the discovery of the relationship between the mass of the central black hole and many properties of the host galaxy, including the mass of the bulge \citep{magorrian98,marconi03,haring04,sani11} and stellar velocity dispersion \citep{ferrarese00,gebhardt00,tremaine02}, indicating co-evolution of black holes and host galaxies \citep[but see also][]{jahnke11}.
More recently, quasar driven outflows have also been observed \citep{feruglio10,cicone12}.
AGN feedback has been implemented in cosmological simulations, which provided a good agreement with the Magorrian relation \citep{springel05} and a better reproduction of the cosmic star formation rate \citep{booth09}.
These simulations used basically the same AGN model that consists of i) black hole formation, ii) evolution, and iii) feedback. 
For the formation, seed black holes with mass $\sim 10^5M_\odot$ are spawned in sufficiently massive dark matter haloes (typically $\sim 10^{10}M_\odot$) not currently possessed of a black hole \citep[e.g.][]{springel05,sijacki07,dimatteo08,booth09}, and the origin of the black holes has not been discussed in detail in the context of cosmological simulations.
We aim to connect black hole formation to the formation of first stars and chemical enrichment.

At the end of the dark age of the Universe, the cosmic dawn was heralded by the birth of the first stars and galaxies.
The nature of these first objects is still far from being well understood.
From the theory of star formation from primordial gas, the first stars are believed to be very massive, with masses of the order of $100M_\odot$, given the limited cooling of molecular hydrogen \citep[e.g.,][]{bromm04}.
This depends on fragmentation in a cosmological minihalo, ionization prior to the onset of gravitational collapse, and the accretion rate from the cloud envelope, and it seems possible to form lower-mass stars ($\sim10-40M_\odot$) from primordial gas in recent numerical simulations \citep[e.g.,][]{yoshida08,stacy10,greif11}.
On the other hand, if the accretion is not suppressed by feedback from the central star, $\sim1000M_\odot$ stars may form \citep{hirano14}.
%
This matches with the observed signatures of the first chemical enrichment. 
The primordial stars with initial masses of $\sim 140-300M_\odot$ explode as pair instability supernovae (PISNe) and leave no remnant, which have large (Si,S)/O ratios \citep[e.g.,][]{nomoto13}.
The observed elemental abundance patterns of extremely metal-poor stars and metal-poor quasar absorption line systems are consistent not with PISNe but with core-collapse supernovae from $\sim 13-50M_\odot$ stars that produce $\sim 10M_\odot$ black holes \citep{ck11dla}.
Stars more massive than $\sim 300M_\odot$ end up as black holes, which can also be seeds of supermassive black holes.


In this paper, we examine whether our scenario can work as the origin of AGN feedback.
In Section \ref{sec:model}, we introduce the code used to perform our simulations, and describe our model for AGN feedback.
We give an overview of the simulations in Section \ref{sec:sims}, and put a constraint on the parameters of the AGN model in Section \ref{sec:ptests}.
In Section \ref{sec:restests} we show some resolution dependence of our results.
Finally, we discuss the implications of our model and draw our conclusions in Sections \ref{sec:disc} and \ref{sec:conc} respectively.



\section{Model}
\label{sec:model}

Our simulation code is based on the smoothed particle hydrodynamics (SPH) code {\small GADGET-3} \citep{springel01gadget,springel05gadget}, which is fully adaptive with individual smoothing lengths and timesteps and uses an entropy-conserving formulation of SPH \citep{springel02}.
Various physical processes associated with the formation and evolution of galaxies are included: radiative cooling, star formation, supernova feedback \citep{ck07}, and black hole physics (this work).

Photo-heating is given by a uniform and evolving UV background radiation \citep{haardt96}.
Radiative cooling is computed using a metallicity-dependent cooling function \citep{sutherland93}, and molecular cooling is not included, and is not important for the temperatures resolved in our simulations.
The star formation criteria are the same as in \citet{katz92}: (1)  converging flow, $(\nabla \cdot \mbox{\boldmath$v$})_i < 0$; (2) rapid  cooling, $t_{\rm cool} < t_{\rm dyn}$; and (3) Jeans unstable gas, $t_{\rm    dyn} < t_{\rm sound}$.
The star formation timescale is taken to be proportional to the dynamical timescale ($t_{\rm sf} \equiv \frac{1}{c_*}t_{\rm dyn}$), where $c_*$ is a star formation timescale parameter which we set to $0.02$ \citep{murante10,scannapieco12}, which gives a better agreement with observations of the cosmic SFR using a Kroupa IMF than $0.1$.
We spawn new star particles as in \citet{ck07}, and follow the evolution of the star particles at every timestep.
Note that this procedure does not use an explicit minimum density threshold to form stars.
The energies are distributed to a constant number of neighbour gas particles $N_{\rm FB}$, which we set to $72$, weighted by an SPH kernel.
The energies of mass loss and supernovae are distributed in purely thermal form, although a fraction of it could, in principle, be distributed in kinetic form as a velocity perturbation to the gas particles \citep{navarro93}.
However, with our scheme of star formation and supernova feedback, we do not need the kinetic feedback or momentum driven winds that are included in some other codes \citep{dave06}.
We also do not apply artificial multi-phase to gas particles that is included in some other codes \citep{scannapieco06}.

The chemical enrichment is computed with the scheme of \citet{ck04}, which does not include the instantaneous recycling approximation.
Different from \citet{ck07}'s simulations, we adopt the initial mass function (IMF) of stars from \citet{kroupa08} and the updated neucleosynthesis yields of \citet{ck11b} for $1-50 M_\odot$. Note that the mass loss from low-mass stars has been included since \citet{ck04}.
The effects of hypernovae are included with a metal-dependent hypernova fraction as in \citet{ck11a}.
The progenitor model of Type Ia supernovae is based on the single degenerate scenario, and taking account of the metallicity effects of white dwarf winds \citep{ck98}, the lifetime distribution functions are calculated as in \citet{ck09}.

\subsection{Black Hole Physics}
\label{sec:agnmodel}

\subsubsection{Seeding black holes}
\label{sec:seeding}

In previous implementations of AGN feedback, a single supermassive black hole is spawned in every dark matter halo of sufficient mass (typically $\sim 10^{10}M_\odot$), identified by regular running of the Friend of Friends routine \citep{springel05,sijacki07,dimatteo08,booth09}.
A gas particle within such a group, which satisfies some criterion (the most gravitationally bound, or the most dense for example), is converted to a black hole, with mass $\sim10^5M_\odot$.
This ensures that galaxies that later form in these haloes contain a black hole.

We adopt a different approach for the formation of seed black holes, motivated by the theory of primordial star formation and the observed signatures of the first chemical enrichment.
Much theoretical work has been undertaken in understanding how the first black holes formed, with the most likely candidates being as the remnants of Population\,{\sc iii} stars or formed by the direct collapse of primordial gas.
For the latter scenario, black holes formed via direct collapse of a massive, low angular momentum gas cloud are likely rare due to the conditions necessary for the collapse to occur.
In order to maintain a high Jeans mass, the gas must not cool efficiently, since $M_\textrm{J}\propto T^{3/2}$.
This may be accomplished by suppressing the formation of H$_\textrm{2}$ with a strong Lyman Werner flux, so that cooling is dominated by atomic hydrogen and the gas is not able to fragment.
Modeling of such situations can produce seed black holes as large as $10^5M_\odot$ \citep{bromm03,koushiappas04,agarwal12}.
For the former scenario, the deaths of Population\,{\sc iii} stars, either as core collapse supernovae ($M_\textrm{star}\lesssim140M_\odot$) or by collapsing directly ($M_\textrm{star}\gtrsim300M_\odot$), could conceivably produce black hole seeds with mass in the range $10\lesssim \ms /M_\odot \lesssim 10^3$ \citep{madau01,schneider02,bromm02}.

Our criteria for forming black holes are therefore as follows: any gas particle satisfying
\begin{equation}\label{eq:bhform1}
	\rho_\textrm{g} > \rc,
\end{equation}
and
\begin{equation}\label{eq:bhform2}
	Z=0,
\end{equation}
where $\rho_\textrm{g}$ and $Z$  are the density and metallicity of the gas particle respectively, and $\rc$ a specified critical density, is converted into a black hole with a constant seed mass $\ms$.
Similar criteria were adopted in \citet{bellovary11} to study the occupation fraction of black holes in simulated haloes.
From the above discussion, we consider the range $10^1\leq \ms /M_\odot \leq 10^5$.
We treat $\rc$ as a free parameter and determine its value by comparing the final state of simulations with observations (see Section \ref{sec:rho}).
We therefore restrict neither the number of black holes that form, nor the locations where they form.

Note that, due to the limited mass resolution, $\ms$ is much smaller than $M_\textrm{gas}$.
As in previous works \citep[e.g.][]{springel05,booth09}, we store the mass of the black hole separately and use $M_\textrm{gas}$ as the dynamical mass until $M_\textrm{BH}$ reaches $M_\textrm{gas}$.
We also note that due to the limited spatial resolution of the simulations presented in this paper (typically a few comoving kpc), it is possible for black holes to `wander' out of galaxies in a few timesteps solely because of numerical effects.
To avoid this artificial effect, at every timestep we calculate the centre of mass of all particles within the smoothing radius of each black hole and reposition the black hole to that location.

\subsubsection{Black hole growth}
\label{sec:bhgrowth}

The seed black holes grow by both gas accretion and mergers with multiple other black holes.
In cosmological hydro-simulations, the spatial resolution is usually $\gtrsim 1h^{-1}$kpc, and thus it is not possible to resolve the detailed physics of accretion near black holes.
It is therefore necessary to use a sub-resolution model for the processes governing the growth of black holes and the feedback from black holes.
We adopt roughly the same model as in many previous works \citep{springel05,sijacki07,dimatteo08,booth09}.
The growth rate due to accretion is taken to be proportional to the Bondi-Hoyle accretion rate \citep{bondi44},
\begin{equation} \label{eq:accrt}
	\dot M_\textrm{acc} = \alpha\frac{4\pi G^2 M^2_\textrm{BH} \rho}{(c^2_\textrm{s}+v^2)^{3/2}},
\end{equation}
where $\rho$ is the gas density local to the black hole, $c_\textrm{s}$ is the sound speed of the gas local to the black hole, $v$ is the relative velocity between the black hole and local gas, and the constant factor $\alpha$ accounts for the finite resolution of the simulations.
In contrast to other works \citep[e.g.,][]{springel05,booth09,barai14}, mass is accreted from gas particles in a continuous fashion, rather than stochastically accreting entire gas particles.
In each timestep $\Delta t$, gas particles neighbouring a black hole have their mass decreased by an amount $\dot M_\textrm{acc}\Delta t$ times their SPH kernel weight.
In principle, this procedure could lead to gas particles with masses much less than the simulation resolution, however the accretion rates are sufficiently low compared to the gas particle mass that this problem does not arise; at $z=0$ in our fiducial simulation, the lowest mass gas particle has 25 per cent of its original mass, and only 3 per cent of gas particles are less massive than initially.

In previous works, values for $\alpha$ as low as 1 \citep{booth09}, or as high as a few hundred \citep{springel05,dimatteo08,khalatyan08} have been used.
We treat $\alpha$ as a free parameter and determine its value by comparing the final state of simulations with observations (see Section \ref{sec:alpha}).
Throughout, we assume that growth is Eddington limited, so that $\dot M_\textrm{acc}\leq \dot M_\textrm{Edd}$ at all times.
The Eddington accretion rate is given by
\begin{equation}\label{eq:edd}
	\dot M_\textrm{Edd} = \frac{4\pi GM_\textrm{BH}m_\textrm{p}}{\er\sigma_\textrm{T}c},
\end{equation}
where $m_\textrm{p}$ is the proton mass, and $\sigma_\textrm{T}$ the Thompson cross section.
Black holes then grow due to accretion at a rate given by
\begin{equation}\label{eq:mbhd}
	\dot M_\textrm{BH} = (1-\er)\times \min(\dot M_\textrm{acc}, \dot M_\textrm{Edd}).
\end{equation}
$\er$ denotes the radiative efficiency of the black hole; we set $\er=0.1$ throughout this paper \citep{shakura73}.

A merger between black holes may occur if the following conditions are met:
\begin{equation}\label{eq:mergevrel}
	v_\textrm{rel} < c_\textrm{s};
\end{equation}
\begin{equation}\label{eq:mergeresl}
	r < \epsilon,
\end{equation}
where $v_\textrm{rel}$ is the relative velocity of the black holes, $r$ their separation, and $\epsilon$ the gravitational softening length of the black holes.
For the condition of equation \eqref{eq:mergevrel}, we use the local sound speed $c_\textrm{s}$ as a measure of the local velocity scale \citep{springel05,dimatteo08}, which precludes black holes undergoing a quick flyby from merging.
We note that \citet{booth09} used $v_\textrm{rel} < \sqrt{GM/\epsilon}$, where $M$ is the mass of the most massive black hole of the pair.
We do not employ this criterion in the current work since the mass of a black hole as calculated by equation \eqref{eq:accrt} can be several orders of magnitude lower than the dynamical mass of the black hole, used to determine its velocity.
Following the merging of black holes, the properties of the primary black hole are updated such that mass and linear momentum are conserved, and it is relocated to the center of mass of the merging particles.
Although the processes of black hole binary hardening is still not well understood \citep{makino04}, and would not be resolved in these simulations of galaxies in any case, the conditions of equations \eqref{eq:mergevrel} and \eqref{eq:mergeresl} cause black holes that would be expected to merge eventually to do so.

\subsubsection{Energy feedback}

In each timestep $\Delta t$, a black hole produces an amount of energy $E_\textrm{FB}$, which is calculated as
\begin{equation}\label{eq:efb}
	E_\textrm{FB} =\er\ef \dot M_\textrm{acc}c^2 \Delta t,
\end{equation}
where $\ef$ denotes the fraction of radiated energy that couples to the gas.
Previous works have used $\ef$ in the range 0.05 \citep{springel05,dimatteo08} to 0.15 \citep{booth09}; we treat $\ef$ as a free parameter and determine its value by comparing the final state of simulations with observations (see Section \ref{sec:epsf}).

$E_\textrm{FB}$ is distributed with the same scheme as for supernova feedback; kernel weighted, to a constant number of neighbouring gas particles.
We adopt $N_{\rm FB}=72$; this value is chosen in order that, at the resolution of our fiducial simulation, gas particles are sufficiently heated that they do not cool too rapidly, but not so much that an excessive mass of gas is expelled from galaxies \citep{ck07}.
The energy contributes only to the internal energy of the gas particles; there is no kinetic component.
We note that the energy coupled to the gas is distributed roughly isotropically at our resolutions, whereas on smaller scales, energy could be injected along jets, as shown in small-scale simulations of AGN \citep[e.g.][]{wagner11}.
With such small scale kinetic feedback, star formation may be enhanced \citep[e.g.][]{silk13}.
A more realistic such model is left to future works.



\section{The Simulations}
\label{sec:sims}

\begin{table*}
\caption{Input parameters and present quantities of the simulations for the parameter study.
(1) Simulation name.
(2) Value of $\alpha$ (see equation \eqref{eq:accrt}).
(3) Value of $\ef$ (see equation \eqref{eq:efb}).
(4) Black hole seed mass.
(5) Value of critical density for black hole formation (see equation \eqref{eq:bhform1}).
(6) Mass of dark matter particles.
(7) Initial mass of gas particles.
(8) Stellar mass fraction $f_* = M_*/(M_*+M_\textrm{g}+M_\textrm{BH})$.
(9) Total number of black holes.
(10) Black hole mass fraction $f_\textrm{BH}=M_\textrm{BH}/(M_\textrm{BH}+M_\textrm{g}+M_*)$.
(11) Average ratio of black hole mass to stellar mass within $R_\textrm{e}$.
(12) Mass of most massive black hole.}
\centering
\begin{sideways}
\begin{tabular}[width=\textwidth]{cccccccccccc}
	\hline
	Name & $\alpha$ & $\ef$ & $\ms$ & $\rc$ & $M_\textrm{DM}$ & $M_\textrm{gas}$ & $f_*$ & $N_\textrm{BH}$ & $f_\textrm{BH}$ & $\left< M_\textrm{BH}/M_{*,R_\textrm{e}} \right>$ & $\max{(M_\textrm{BH})}$\\
	& & & ($h^{-1}$M$_\odot$) & ($h^2$m$_\textrm{H}\,$cm$^{-3}$) & ($\times10^7\,h^{-1}\textrm{M}_\odot$) & ($\times10^7\,h^{-1}\textrm{M}_\odot$) & & & ($\times10^{-5}$) & ($\times10^{-3}$) & ($\times10^7\,h^{-1}\textrm{M}_\odot$)\\
	(1) & (2) & (3) & (4) & (5) & (6) & (7) & (8) & (9) & (10) & (11) & (12)\\
	\hline
	\hline
	F & 1 & 0.25 & $10^3$ & 0.1 & 7.34 & 1.44 & 0.053 & 206 & 1.36 & 0.7 & 7.4\\
	\hline
	A1 & 10 & 0.25 & $10^3$ & 0.1 & 7.34 & 1.44 & 0.039 & 195 & 1.98 & 3.2 & 9.4\\
	A2 & 50 & 0.25 & $10^3$ & 0.1 & 7.34 & 1.44 & 0.030 & 212 & 2.26 & 7.1 & 9.1\\
	A3 & 100 & 0.25 & $10^3$ & 0.1 & 7.34 & 1.44 & 0.026 & 207 & 2.23 & 5.0 & 10.9\\
	\hline
	E1 & 1 & 0.01 & $10^3$ & 0.1 & 7.34 & 1.44 & 0.049 & 201 & 22.53 & 12.5 & 126\\
	E2 & 1 & 0.05 & $10^3$ & 0.1 & 7.34 & 1.44 & 0.050 & 198 & 5.58 & 2.4 & 34.1\\
	E3 & 1 & 0.15 & $10^3$ & 0.1 & 7.34 & 1.44 & 0.054 & 205 & 2.12 & 0.9 & 11.8\\
	E4 & 1 & 0.50 & $10^3$ & 0.1 & 7.34 & 1.44 & 0.054 & 205 & 0.82 & 0.4 & 4.1\\
	\hline
	M1 & 1 & 0.25 & $10^1$ & 0.1 & 7.34 & 1.44 & 0.071 & 210 & 3e-3  & 3e-3 & 9e-3\\
	M2 & 1 & 0.25 & $10^2$ & 0.1 & 7.34 & 1.44 & 0.069 & 200 & 0.15 & 5e-2 & 1.1\\
	M3 & 1 & 0.25 & $10^4$ & 0.1 & 7.34 & 1.44 & 0.039 & 196 & 4.56 & 8.7 & 18.0\\
	M4 & 1 & 0.25 & $10^5$ & 0.1 & 7.34 & 1.44 & 0.018 & 207 & 24.77 & 94.3 & 89.7\\
	\hline
	R1 & 1 & 0.25 & $10^3$ & 0.5 & 7.34 & 1.44 & 0.072 & 126 & 0.60 & 3.2 & 2.0\\
	R2 & 1 & 0.25 & $10^3$ & 1.0 & 7.34 & 1.44 & 0.075 & 108 & 0.40 & 0.3 & 1.4\\
	\hline
\end{tabular}
\end{sideways}
\label{tab:ptest}
\end{table*}

\begin{figure*}
\centering
\includegraphics[width=\textwidth,keepaspectratio]{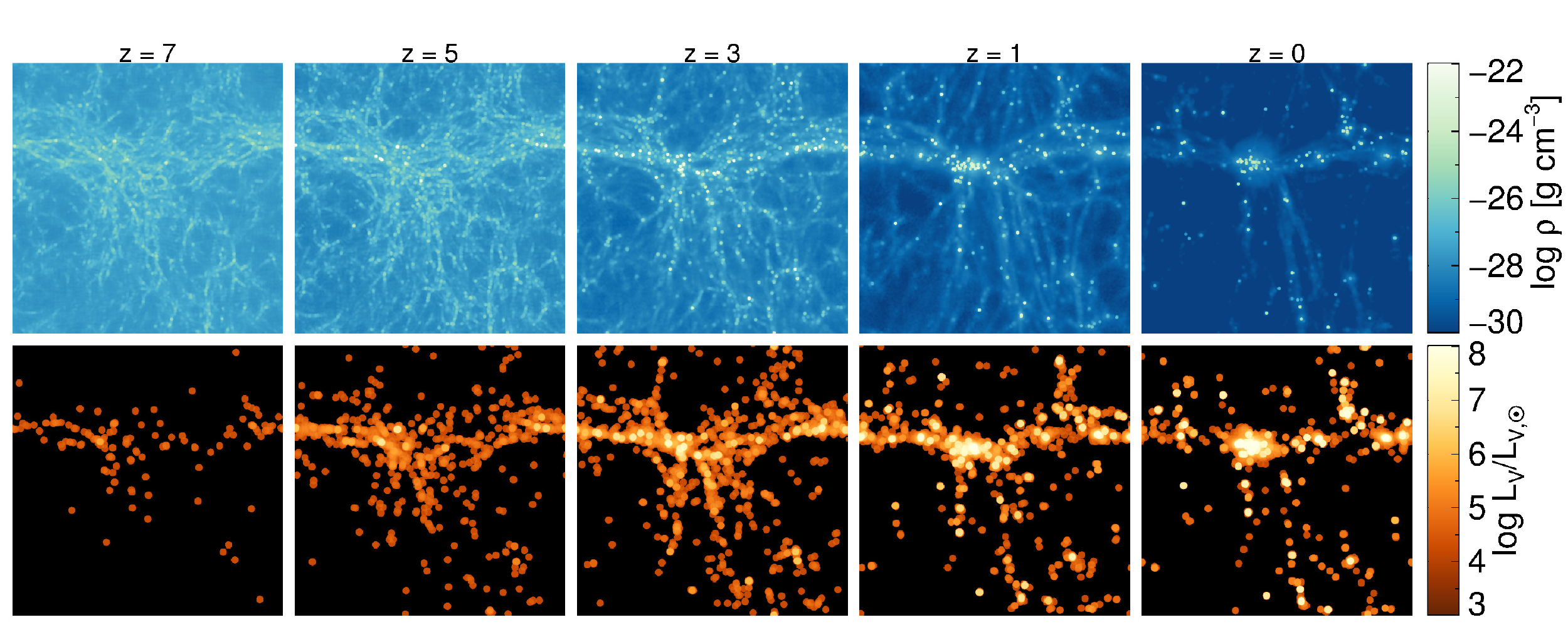}
\caption{The time evolution of our fiducial simulation, F, in a periodic box 10 $h^{-1}$Mpc on a side.
We show projected gas density on the upper row, and V band luminosity on the lower row.}
\label{fig:evol}
\end{figure*}

\begin{figure}
\centering
\includegraphics[width=0.47\textwidth,keepaspectratio]{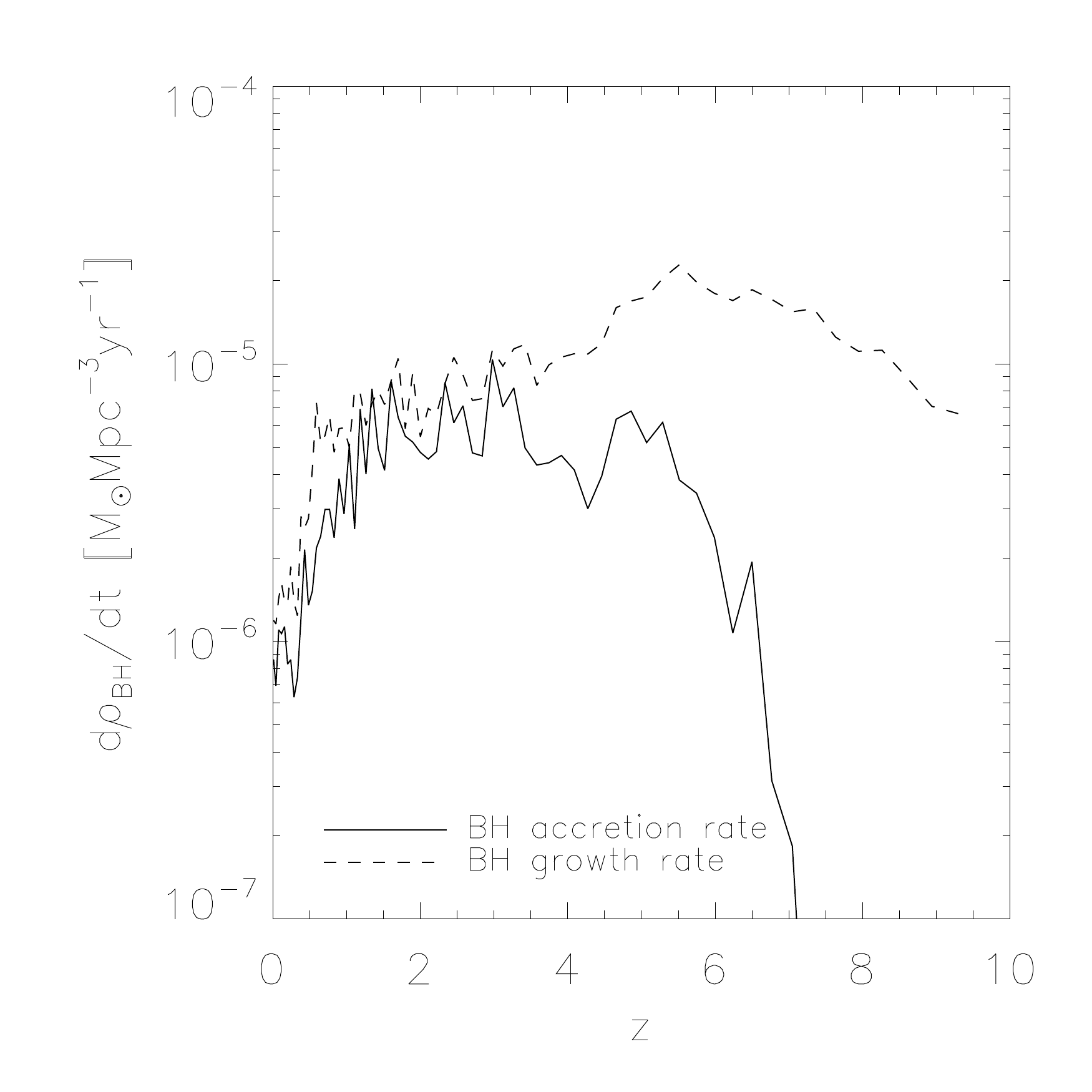}
\caption{Cosmic accretion rate density (solid line) and growth rate density (including black hole formation, dashed line) of our fiducial simulation, F, with redshift.}
\label{fig:acc}
\end{figure}

\begin{figure*}
\centering
\includegraphics[width=\textwidth,keepaspectratio]{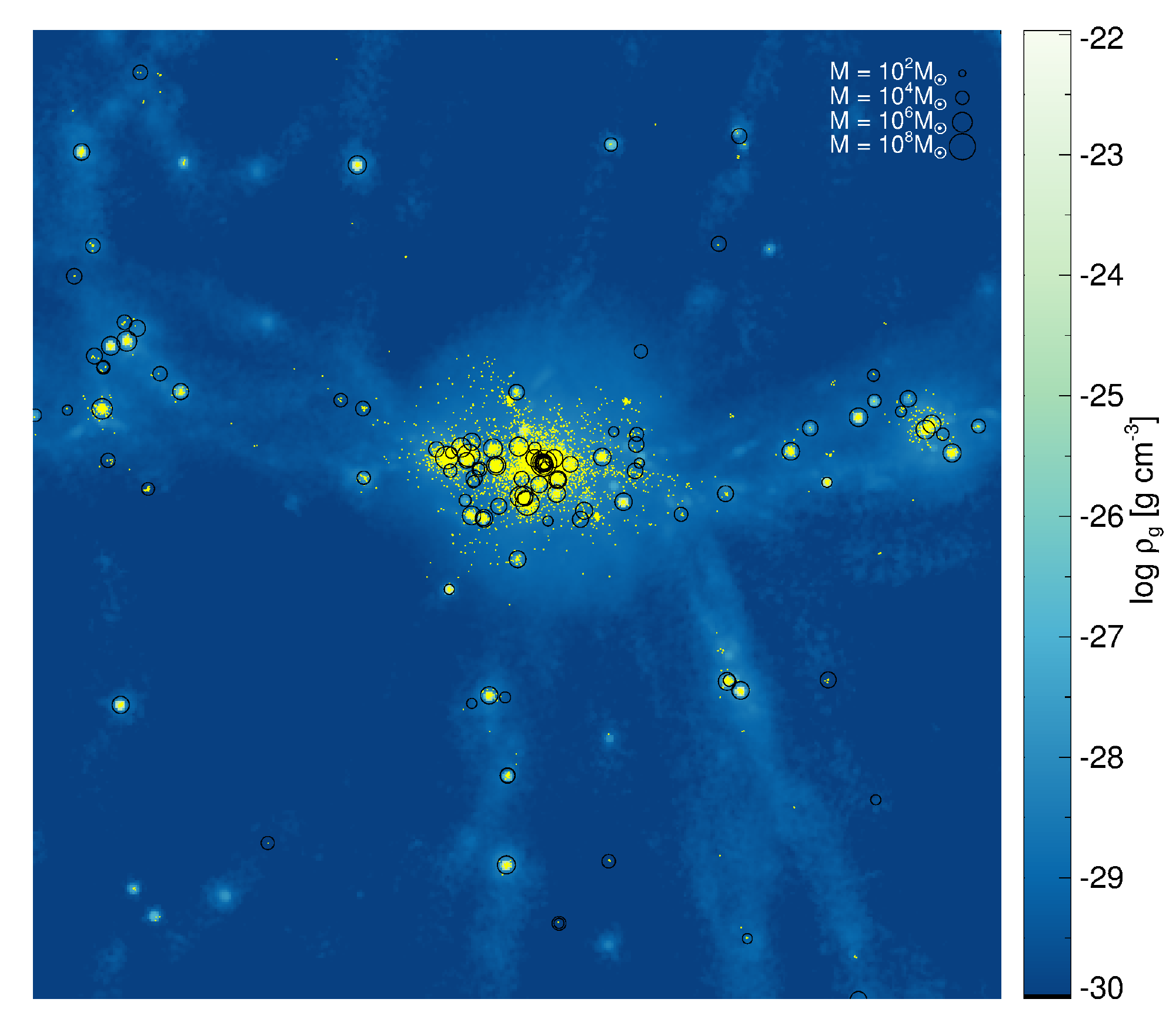}
\caption{The distribution of black holes (black circles, whose radii are proportional to $\log M_\textrm{BH}$) and star particles (yellow dots) in a $6\times 6\times 10$ ($h^{-1}$Mpc)$^3$ volume at $z=0$ in our fiducial simulation.
We also show the projected gas density (blue).}
\label{fig:densbh}
\end{figure*}

We employ a WMAP-9 $\Lambda$CDM cosmology \citep{wmap9} with $h=0.70$, $\Omega_\textrm{m}=0.28$, $\Omega_\Lambda=0.72$, $\Omega_\textrm{b}=0.046$, and $\sigma_8=0.82$.
The simulations presented in Section \ref{sec:ptests} are run at the same resolution with identical initial conditions, in a periodic, comoving, cubic volume 10 $h^{-1}$ Mpc on a side.
This initial condition is chosen to have a central concentration at $z=0$, and is different from \citet{ck07}, which corresponds to typical field of the Universe.
The parameters used in each simulation are listed in Table \ref{tab:ptest}, as well as $z=0$ properties of the simulated volumes.
The initial conditions contain 96$^3$ particles of each of dark matter and gas, with masses $M_\textrm{DM}=7.3\times 10^7\,h^{-1}M_\odot$ and $M_\textrm{gas}=1.4\times 10^7\,h^{-1}M_\odot$; by $z=0$ there are approximately 5 per cent more particles in total due to the formation of star particles.
We use a gravitational softening length of $\epsilon_\textrm{gas}=1.125\,h^{-1}$ kpc.

The redshift evolution of our fiducial simulation (see Section \ref{sec:ptests}) is shown in Fig. \ref{fig:evol}.
At high redshift, a rich filamentary structure exists, to which star and black hole formation is mostly confined.
The first black hole forms at $z\sim16$ (at the resolution of these simulations) in the region that will subsequently collapse to form the largest cluster in the simulation box.
This first black hole grows and provides thermal feedback to neighbouring gas particles, as described in Section \ref{sec:agnmodel}.
Heating by AGN feedback results in the delay of the formation of the first star until $z\sim11$.
Following the hierarchical clustering of dark haloes, cold gas falls along the filaments, which boosts the formation of black holes and stars.
Chemical enrichment from the previous generation of stars suppresses the formation of black holes but enhances star formation.
As a result, the SFR peaks at $z\sim 2$ (see Figs \ref{fig:alpha}-\ref{fig:rhosfr} in Section \ref{sec:ptests}), coincident with a broader peak in the cosmic black hole accretion rate.
The black hole number density peaks earlier, at $z\sim5.5$, after which mergers reduce their total number.
Fig. \ref{fig:acc} shows the cosmic black hole accretion rate (solid line) and growth rate including black hole formation (dashed line), for our fiducial simulation.
Gas accretion only becomes important in the growth of black holes from $z\sim4$.
By $z=0$, the accretion rate has fallen to $\sim10^{-6}M_\odot\textrm{Mpc}^{-3}\textrm{yr}^{-1}$, consistent with the observational estimate of $1.3\times10^{-6}M_\odot\textrm{Mpc}^{-3}\textrm{yr}^{-1}$ from radio luminosities \citep{smolcic09}.

Due to the supernova and AGN feedback, galactic winds are generated, which eject interstellar gas into the intergalactic medium.
The largest concentration of stars and black holes is seen in a central cluster of galaxies, with a present total mass of $\sim 10^{13}\,h^{-1}M_\odot$, and which hosts the most massive black hole in the simulation volume with a mass $\sim10^8\,h^{-1}M_\odot$.
In Fig. \ref{fig:densbh} we show the $z=0$ spatial distribution of stars and black holes in a $6\times 6\times 10$ ($h^{-1}$Mpc)$^3$ volume around the largest cluster, as well as the projected gas density.
Star and black hole particles are found along filaments, coinciding with local peaks in gas density.
Although our description of seeding of black holes does not ensure the co-evolution between black holes and galaxies, larger black holes tend to be located in larger galaxies (see Figs. \ref{fig:alphambhsig}, \ref{fig:epsmbhsig}, \ref{fig:mseedmbhsig}, and \ref{fig:rhombhsig}).

Galaxies are identified using a Friend of Friends algorithm, in common with previous works \citep[e.g.][]{springel01,ck07,booth09}.
The algorithm identifies groups of dark matter particles, using a comoving linking length of 0.02 times the mean dark matter particle separation.
Gas, star, and black hole particles are then associated with their nearest dark matter particle, and join that particle's group.
A catalogue is then made of all groups whose total mass is at least $32M_\textrm{DM}$.
For the analysis of properties of a galaxy, only those groups containing at least 50 star particles are kept.
In the fiducial simulation at $z=0$, 99 groups are found by the friend of friends algorithm, 76 are classified as galaxies, corresponding to $10^9\lesssim M_*/M_\odot\lesssim5\times10^{11}$, and 60 of these contain one or more black holes.
4 galaxies host multiple black holes: the most massive galaxy contains 6, while 3 others have 2 black holes each.
In all cases, one of the black holes is much more massive than the others; in the most extreme case, that of the most massive galaxy, the most massive black hole contains 99.5 per cent of the combined mass of the 6.


The $z=0$ stellar and black hole mass fractions, number of black holes, and the mass of the largest black hole in the simulations of Section \ref{sec:ptests} are given in Table \ref{tab:ptest}.
Larger $\alpha$ and $\ms$, and smaller $\rc$, results in stronger feedback, which leads to a greater reduction in star formation, and more massive black holes.
Larger $\ef$ results in more massive black holes, but does not affect star formation history.
Without AGN feedback, the $z=0$ stellar mass fraction, $f_*=M_*/(M_*+M_\textrm{g}+M_\textrm{BH})$, is 0.075, which is consistent with the result in \citet{ck07}.
With AGN feedback included, $f_*$ is reduced to 0.053 in our fiducial simulation, which is more comparable with the observational estimate of $\sim0.046$ \citep[e.g.][]{fukugita04}.
Observationally, the black hole to bulge mass ratio in galaxies is found to be $M_\textrm{BH}/M_\textrm{bulge}\sim2\times10^{-3}$ \citep[e.g.][]{marconi03}.
We find $\left<M_\textrm{BH}/M_*(<R_\textrm{e})\right>=0.7\times10^{-3}$, with $R_\textrm{e}$ the effective radius of a galaxy (see Section \ref{sec:ptests} for more details).
The small discrepancy comes from the difference between $M_\textrm{bulge}$ and $M_*$; the bulge mass of late type galaxies tends to be smaller than $M_*(<R_\textrm{e})$.


\section{Parameter Study}
\label{sec:ptests}

In this Section, we present the results of a suite of simulations, each with different values of the free parameters of the model described in Section \ref{sec:model}.
We then determine a `best' set of parameters, such that the simulation using these parameters produce results that most closely match observations.
For the observational constraints, we use cosmic SFR history, size--velocity dispersion relations of galaxies \citep{trujillo11}, and the $M_\textrm{BH}-\sigma$ relation \citep{kormendyho13}. 
For the cosmic SFR history, observational data are taken from \citet{bouwens11}, \citet{karim11}, \citet{cucciati12}, \citet{oesch12}, \citet{burgarella13}, \citet{gunawardhana13}, and \citet{sobral13}, adjusted to the \citet{kroupa08} IMF \citep{bernardi10}.
These values are systematically lower than \citet{hopkins06}.
Values derived from H$\alpha$ measurements may be underestimated due to the bivariate selection process \citep{gunawardhana13}, and those from UV measurements are sensitive to dust extinction.

For the size--velocity dispersion relation, we fit a S\'ersic profile to the surface density of the galaxies, and define the effective radius, $R_\textrm{e}$, as that which encloses half the stellar mass.
Note that $R_\textrm{e}$ tends to be large due to the spatial resolution of our simulations.
The velocity dispersion is measured within the central $R_\textrm{e}$ of galaxies.
We compare the $M_\textrm{BH}-\sigma$ relations obtained from our simulations to the one presented in \citet{kormendyho13}, who find $\log(M_\textrm{BH}/10^9M_\odot)=-(0.510\pm 0.049) + (4.377\pm 0.290)\log(\sigma/200\,\textrm{km s}^{-1})$.
This is in keeping with the findings of other studies \citep{tremaine02,hu08,gultekin09,greene10}, although some have found steeper slopes \citep[e.g.][]{graham13}.
At least some of this disparity is likely due to the different relation observed for bulges that have formed by mergers and those that have not \citep{hu08,beifiori12,kormendyho13}.
We do not make such a distinction in this work, and compare the galaxy population as a whole to observations.


The values of the best set of parameters are $\alpha=1$, $\ef=0.25$, $\ms=10^3\,h^{-1}M_\odot$, $\rc=0.1\,h^2m_\textrm{H}\,\textrm{cm}^{-3}$.
We use this as our fiducial set of parameters (simulation F) in the following subsections.
In each subsection, we fix the values of all parameters, except that which is being examined, to their fiducial values.

We first perform this parameter study for a wide range of parameter space at moderate resolution, before homing in on the values that comprise our best set.
We then investigate a smaller region of parameter space around these values at higher resolution (see Table \ref{tab:ptest} for details).
The results of these higher resolution tests, and the effects of each of the parameters on the evolution of black holes and galaxies, are presented below, but similar trends are seen for all the tests we have performed.

\subsection{Accretion Rate Modifier}
\label{sec:alpha}

$\alpha$ controls the rate of gas accretion onto black holes via equation \eqref{eq:accrt}, and therefore influences the amount of feedback energy they produce via equation \eqref{eq:efb}.
As discussed in Section \ref{sec:model}, $\alpha$ can be larger than 1, and we apply $\alpha =1, 10, 50, 100$ for the simulations F and A1-3 (see Table \ref{tab:ptest}).

\begin{figure*}
	\begin{subfigure}{0\textwidth}\caption{}\label{fig:alphasfr}\end{subfigure}
	\begin{subfigure}{0\textwidth}\caption{}\label{fig:alphambhsig}\end{subfigure}
	\begin{subfigure}{0\textwidth}\caption{}\label{fig:alphambhz}\end{subfigure}
	\begin{subfigure}{0\textwidth}\caption{}\label{fig:alpharesig}\end{subfigure}
	\begin{subfigure}{0\textwidth}\caption{}\label{fig:alphabmstr}\end{subfigure}
	\begin{subfigure}{0\textwidth}\caption{}\label{fig:alphadensz}\end{subfigure}
	\centering
	\includegraphics[width=0.785\textwidth,keepaspectratio]{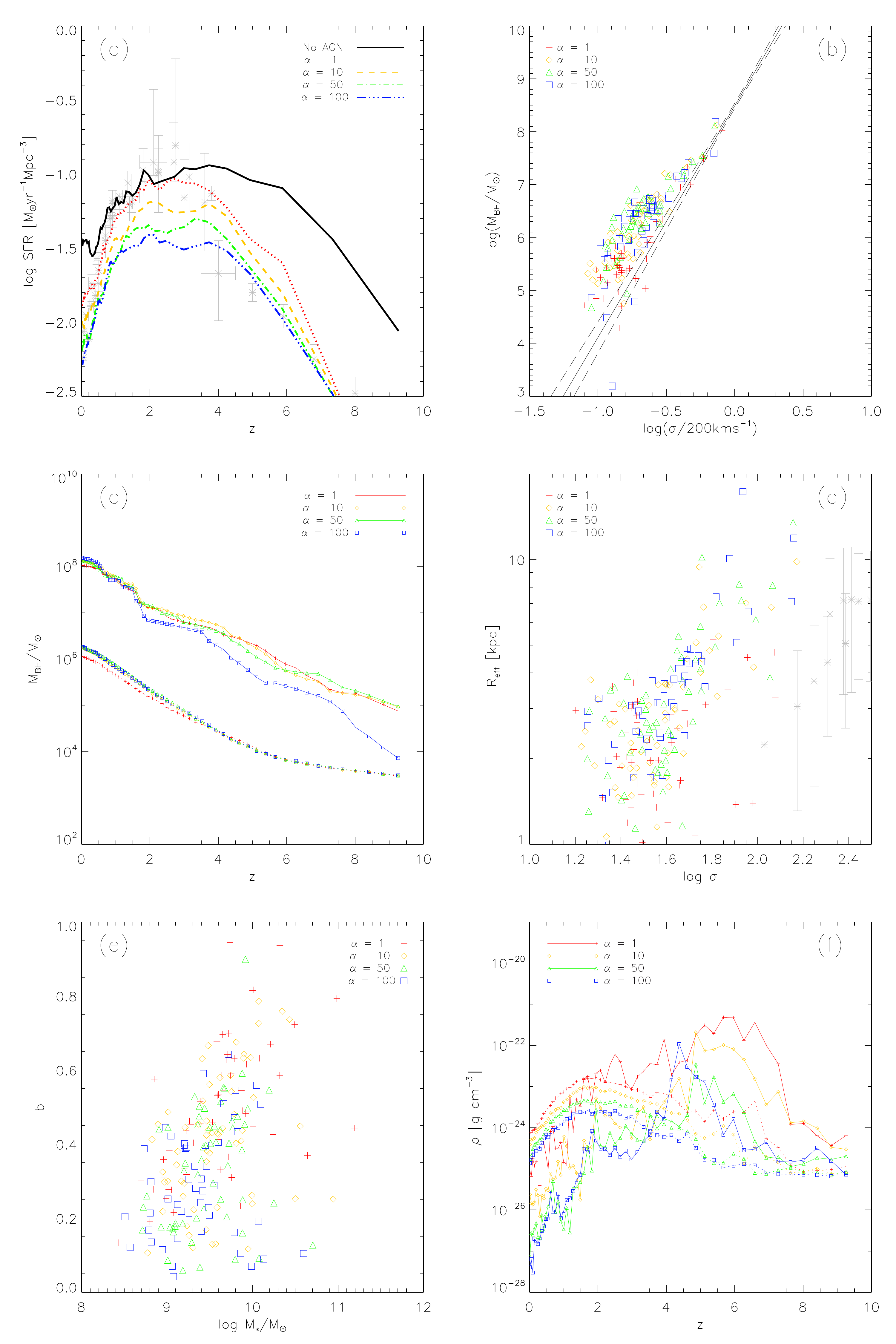}
	\caption{The influence of $\alpha$ on local and global quantities.
(a) Cosmic star formation rate history.
Observational data (grey) is taken from \citet{bouwens11}, \citet{karim11}, \citet{cucciati12}, \citet{oesch12}, \citet{burgarella13}, \citet{gunawardhana13}, and \citet{sobral13}.
(b) Present day $M_\textrm{BH}-\sigma$ relation.
The solid and dashed lines are for the observed relation and 1$\sigma$ width \citep{kormendyho13}, respectively.
(c) Black hole mass as a function of redshift.
Solid lines are for the most massive black hole at $z=0$, and the dotted lines for the average black hole mass in the simulation box.
(d) Present relationship between galaxy effective radius, and galaxy velocity dispersion for all simulated galaxies.
Observational data (grey) is taken from \citet{trujillo11}.
(e) Present relationship between baryon fraction, $b=(M_\textrm{gas}+M_*)/M_\textrm{tot}$, and central stellar mass for all simulated galaxies.
Both quantities are measured in a spherical region at the centre of galaxies of radius $2R_\textrm{e}$.
(f) Gas density local to black holes as a function of redshift (i.e. $\rho$ in equation \eqref{eq:accrt}).
Solid lines are for the most massive black hole at $z=0$, and the dotted lines for the average gas density around all black holes in the simulation box.}
	\label{fig:alpha}
\end{figure*}

One of the clearest effects of changing the value of $\alpha$ is apparent in Fig. \ref{fig:alphasfr}, which shows cosmic SFR as a function of redshift.
All curves show a peak in SFR at $z\sim2$, in agreement with both the simulation without AGN feedback, and observations \citep{hopkins06,karim11,cucciati12,burgarella13,gunawardhana13,sobral13}.
Of the simulations with AGN feedback, all but the $\alpha=1$ case show suppressed star formation with respect to the simulation without AGN across all of cosmic time.
For the fiducial simulation, with $\alpha=1$, the peak is almost the same as in the simulation without AGN, and is suppressed at both earlier and later times, which is in better agreement with the observed low SFRs at $z>6$ \citep{bunker10,bouwens11} than the simulation without AGN.
There is a clear trend for simulations with higher $\alpha$ to have more heavily suppressed star formation.
Na\"{i}vely, one might assume that this is due to $\alpha$ enhancing the accretion rate leading to a greater amount of feedback energy so that some gas particles no longer satisfy the star formation criterion of rapid cooling.
However, this is not the case; the effects of the value of $\alpha$ on the growth of black holes is shown in Fig. \ref{fig:alphambhz}.
The solid lines show the growth of the most massive present day black hole, while the dotted lines show the mean black hole mass of the entire volume.
It is surprising to note that the growth history of both the most massive black hole, and the mean black hole mass, seem to be insensitive to the value of $\alpha$, given that its purpose was to modify the accretion rate.
If black hole growth is self regulated, then the amount of feedback energy released by a black hole over its lifetime should be independent of $\alpha$.
This then implies, from equation \eqref{eq:efb}, that the mass a black hole accretes should also be independent of $\alpha$.
This self-regulation will be discussed further in Section \ref{sec:disc}.

The independence of black hole mass on $\alpha$ is also apparent from Fig. \ref{fig:alphambhsig}, which shows the $z=0$ $M_\textrm{BH}-\sigma$ relation for the simulated galaxies, with the observational relation and scatter represented by the solid and dashed lines respectively.
The fiducial simulation is in good agreement with the observed relation, at the high mass end in particular.

In contrast to the lack of strong dependence of the evolution of black holes, the value of $\alpha$ has a large impact on the properties of their host galaxies.
For example, Fig. \ref{fig:alpharesig}, which shows the $z=0$ relation between galaxy effective radius and velocity dispersion, indicates that there is a trend for galaxies in simulations with large $\alpha$ to have larger effective radius.
This is due to the fact that larger $\alpha$ suppresses star formation from cooling gas near the centre of galaxies to a greater extent, which results in a more extended distribution of stars at a given galaxy mass.
This is the same effect as shown for the dependence of the star formation parameter on the galaxy size in \citet{ck05}.
The fiducial simulation compares well with observations, while those with larger $\alpha$ lie above the observed trend.

As well as the stellar component, the gas component of the simulated galaxies is also affected.
We see a trend in Fig. \ref{fig:alphabmstr} for galaxies formed in simulations with larger values of $\alpha$ to have both lower baryon fractions and lower total stellar mass, within two effective radii.
With $\alpha=1$, the baryon fraction reaches $b\sim1$ in massive galaxies, which is an improvement from \citep{ck05}, and comparable to observational estimates of elliptical galaxies \citep[e.g.,][]{gerhard01}.
This is because gas is removed from the central region of galaxies more efficiently when larger $\alpha$ is used.
This is further evidenced by Fig. \ref{fig:alphadensz}, which shows the gas density local to the most massive black hole (solid lines), and the average gas density around black holes (dotted lines).
There is a clear trend for larger $\alpha$ to lead to a reduction in gas density near black holes.
It is therefore the effect of $\alpha$ on gas density, rather than the amount of feedback energy, that causes the reduction in star formation seen above, by increasing the dynamical time of the gas.

\subsection{Feedback Efficiency}
\label{sec:epsf}

\begin{figure*}
	\begin{subfigure}{0\textwidth}\caption{}\label{fig:epssfr}\end{subfigure}
	\begin{subfigure}{0\textwidth}\caption{}\label{fig:epsmbhsig}\end{subfigure}
	\begin{subfigure}{0\textwidth}\caption{}\label{fig:epsmbhz}\end{subfigure}
	\begin{subfigure}{0\textwidth}\caption{}\label{fig:epsresig}\end{subfigure}
	\begin{subfigure}{0\textwidth}\caption{}\label{fig:epsbmstr}\end{subfigure}
	\begin{subfigure}{0\textwidth}\caption{}\label{fig:epsdensz}\end{subfigure}
	\centering
	\includegraphics[width=0.88\textwidth,keepaspectratio]{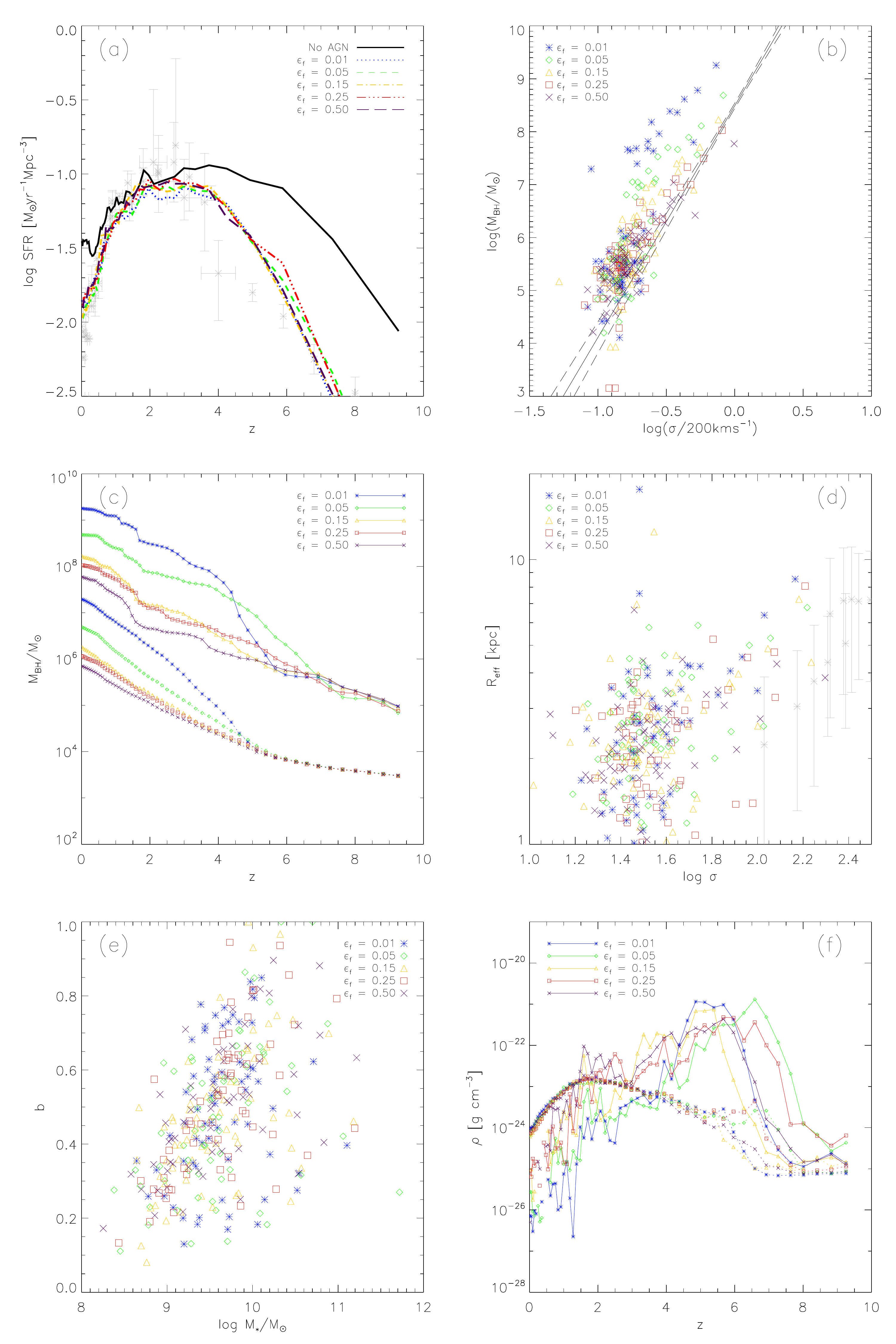}
	\caption{Same as Fig. \ref{fig:alpha}, but for varying $\ef$, and $\alpha=1$, $\ms=10^3\,h^{-1}M_\odot$, $\rc=0.1\,h^2m_\textrm{H}\,\textrm{cm}^{-3}$.}
	\label{fig:eps}
\end{figure*}

$\ef$ determines the fraction of energy produced by a black hole that couples to neighbouring gas particles (see equation \eqref{eq:efb}).
$\ef$ must be less than 1, and we apply $\ef = 0.01,0.05,0.15,0.25,0.50$ for the simulations F and E1-4 (see Table \ref{tab:ptest}).
It is perhaps surprising then that Fig. \ref{fig:epssfr} shows that the value of $\ef$ seems to have no bearing on the cosmic SRF history, other than to suppress star formation with respect to the simulation without AGN feedback.
However, Fig. \ref{fig:epsmbhz} shows that the value of $\ef$ has an effect on the mass of black holes for $z<6$, with lower values allowing for more massive black holes.
For self-regulated black hole growth, the total amount of feedback energy produced by a black hole over its lifetime should be independent of the value of $\ef$, explaining why no variation is seen in star formation histories.
However, from equation \eqref{eq:efb}, this means that the total mass a black hole accretes over its lifetime is smaller for larger $\ef$, as seen.
This self-regulation will be discussed further in Section \ref{sec:disc}.

This trend is also seen in the $M_\textrm{BH}-\sigma$ relation in Fig. \ref{fig:epsmbhsig}.
Lower values of $\ef$ make black holes more massive at a given galaxy mass; with low $\ef$, black holes grow above the observed relation, but with higher values black holes lie on the observed relation.
We choose $\ef=0.25$ as our fiducial value because the high mass end of the relation best matches the observed relation in this case.

Unlike $\alpha$, the value of $\ef$ has no discernible effect on the internal structure of galaxies.
There is no trend with $\ef$ in the relation between effective radius and velocity dispersion at $z=0$ \ref{fig:epsresig}.
The value of $\ef$ has no global effect on the baryonic processes occurring within galaxies.
No significant trend is seen in both Fig. \ref{fig:epsbmstr}, which shows the baryon fraction of galaxies as a function of stellar mass, and Fig. \ref{fig:epsdensz}, which shows the gas density local to black holes as a function of redshift. 
This is another consequence of black hole growth being self-regulated - a black hole removes the same mass of gas from its galaxy, and does not affect star formation, regardless of the value of $\ef$, and so we see no change in the properties of the galaxy population.

\subsection{Seed Black Hole Mass}
\label{sec:mseed}

\begin{figure*}
	\begin{subfigure}{0\textwidth}\caption{}\label{fig:mseedsfr}\end{subfigure}
	\begin{subfigure}{0\textwidth}\caption{}\label{fig:mseedmbhsig}\end{subfigure}
	\begin{subfigure}{0\textwidth}\caption{}\label{fig:mseedmbhz}\end{subfigure}
	\begin{subfigure}{0\textwidth}\caption{}\label{fig:mseedresig}\end{subfigure}
	\begin{subfigure}{0\textwidth}\caption{}\label{fig:mseedbmstr}\end{subfigure}
	\begin{subfigure}{0\textwidth}\caption{}\label{fig:mseeddensz}\end{subfigure}
	\centering
	\includegraphics[width=0.88\textwidth,keepaspectratio]{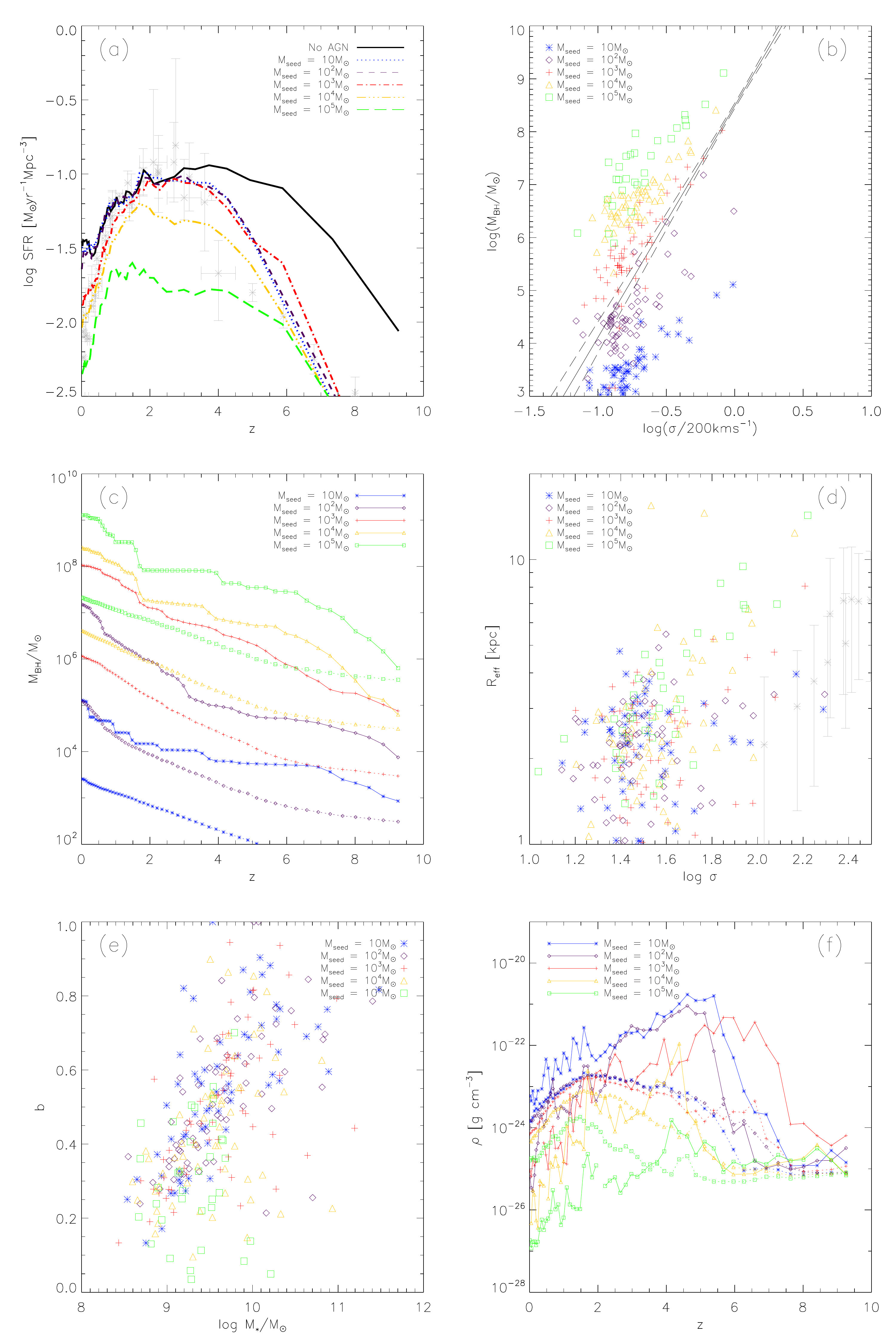}
	\caption{Same as Fig. \ref{fig:alpha}, but for varying $\ms$, and $\alpha=1$, $\ef=0.25$, $\rc=0.1\,h^2m_\textrm{H}\,\textrm{cm}^{-3}$.}
	\label{fig:mseed}
\end{figure*}

The mass with which a black hole is seeded, $\ms$, has a large impact on the subsequent evolution of the black hole itself and its host galaxy.
This is to be expected since, from equation \eqref{eq:accrt}, the accretion rate (and hence also the feedback energy) is proportional to the square of the mass.
We apply $\ms/M_\odot = 10^1,10^2,10^3,10^4,10^5\,h^{-1}$ for simulations F and M1-4 (see Table \ref{tab:ptest}).

This is exemplified by Fig. \ref{fig:mseedsfr}, which shows the SFR as a function of redshift for different values of seed mass.
There is a general trend that a larger seed mass suppresses star formation more efficiently across all time.
This is easily explained by noting that a larger seed mass allows for the growth of more massive black holes, as indicated by Fig. \ref{fig:mseedmbhz}.

We can see again, in Fig. \ref{fig:mseedmbhsig}, how the final mass of the black holes is affected by their seed mass.
With the seed mass as high as $\ms=10^5\,h^{-1}M_\odot$, black holes grow several orders of magnitude above the observed $M_\textrm{BH}-\sigma$ relation.
Only in the fiducial simulation, using a seed mass of $10^3M_\odot$, do the black holes grow onto the observed relation.
We discuss this result more in Section \ref{sec:disc}, connecting it to the astrophysical origin of black hole seeds.

\begin{figure*}
	\begin{subfigure}{0\textwidth}\caption{}\label{fig:rhosfr}\end{subfigure}
	\begin{subfigure}{0\textwidth}\caption{}\label{fig:rhombhsig}\end{subfigure}
	\begin{subfigure}{0\textwidth}\caption{}\label{fig:rhombhz}\end{subfigure}
	\begin{subfigure}{0\textwidth}\caption{}\label{fig:rhoresig}\end{subfigure}
	\begin{subfigure}{0\textwidth}\caption{}\label{fig:rhobmstr}\end{subfigure}
	\begin{subfigure}{0\textwidth}\caption{}\label{fig:rhodensz}\end{subfigure}
	\centering
	\includegraphics[width=0.88\textwidth,keepaspectratio]{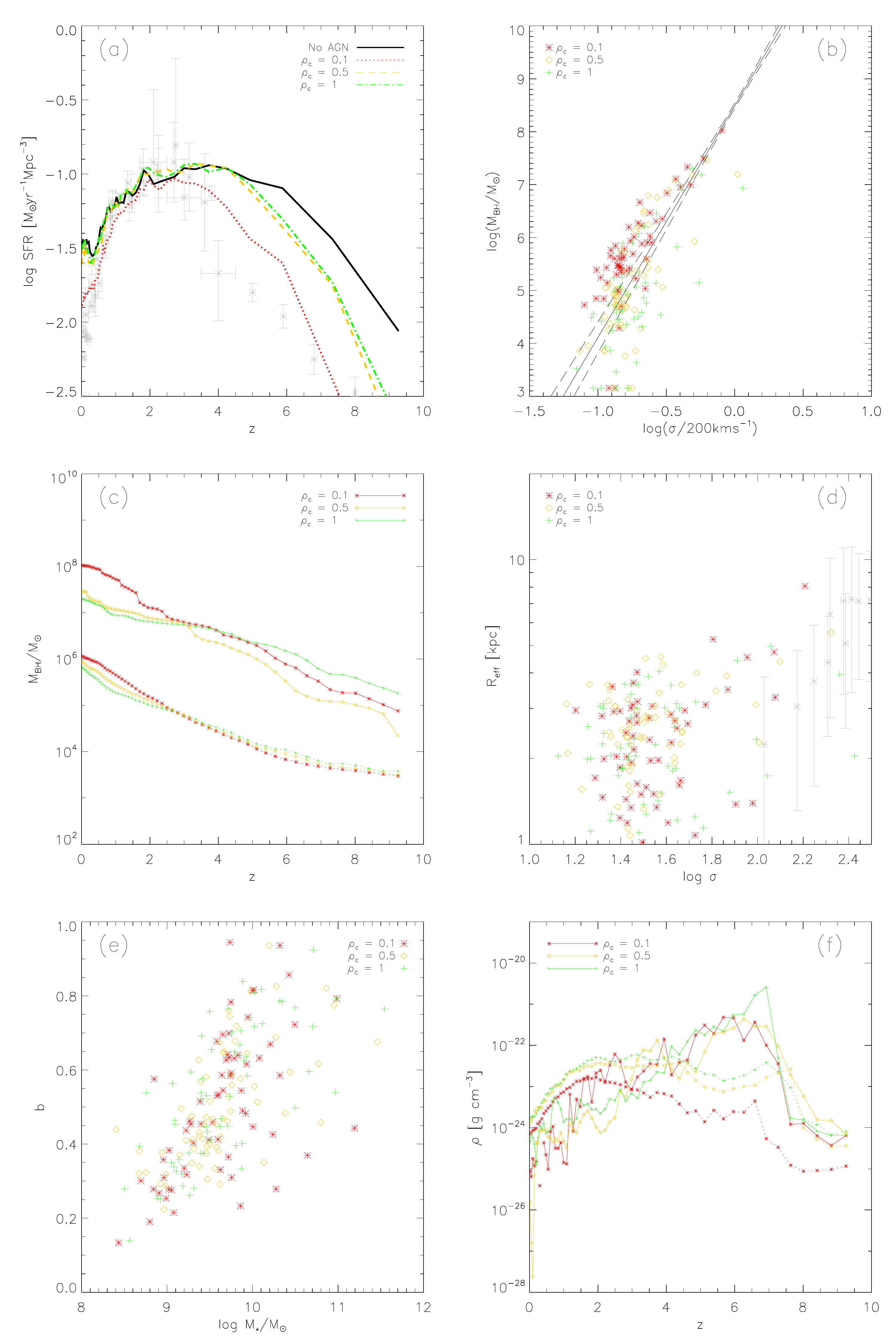}
	\caption{Same as Fig. \ref{fig:alpha}, but for varying $\rc$, and $\alpha=1$, $\ef=0.25$, $\ms=10^3\,h^{-1}M_\odot$.}
	\label{fig:rho}
\end{figure*}

In common with $\alpha$, the adopted value of $\ms$ affects the internal structure of galaxies.
The $z=0$ relationship between galaxy effective radius and velocity dispersion is shown in Fig. \ref{fig:mseedresig}, in which a trend can be seen for larger seed masses to produce galaxies with larger effective radii.
This is due to the fact that more massive black holes quench star formation to a greater extent at the center of galaxies, producing shallower intensity profiles.
Galaxies formed in simulations with higher seed masses are seen to have a lower baryon fraction, as well as a less massive stellar component, as seen in  Fig. \ref{fig:mseedbmstr}, which shows the baryon fraction as function of stellar mass within two effective radii of galaxies.
This is because more massive black holes are able to remove gas  more easily from the centres of galaxies.
This can be clearly seen in Fig. \ref{fig:mseeddensz}, which shows the gas density local to the most massive black hole of each simulation (solid lines), and the average gas density around all black holes in the simulation box (dotted lines).

\subsection{Critical Density}
\label{sec:rho}

$\rc$ is the critical density for black hole formation, which directly determines when black holes form, and how common they are.
It is from these two factors that the subsequent effects of the value of $\rc$ stem.
We apply $\rc=0.1,0.5,1.0 \,h^2m_\textrm{H}\,\textrm{cm}^{-3}$.
Lower $\rc$ allows black holes to form earlier, and at more locations throughout the simulation volume.
This then gives the black hole seeds more time to grow and affect their surroundings, although, by definition, starting from a region with lower gas density.
With $\rc=0.1\,h^2m_\textrm{H}\,\textrm{cm}^{-3}$, black holes can form as early as $z\sim16$, whereas with $\rc = 1\,h^2m_\textrm{H}\textrm{cm}^{-3}$, black holes start forming from $z\sim13.6$, a difference of about 65 Myr.

The effect of this delay on SFR is apparent in Fig. \ref{fig:rhosfr}, in which we see that star formation at high redshift is reduced to a greater extent when $\rc$ is set to a lower value.
For $\rc \ge 0.5\,h^2m_\textrm{H}\,\textrm{cm}^{-3}$, the cosmic star formation history at $z\gtrsim4$ is almost the same, regardless of the value of $\rc$ used; the black holes form late enough, and in small enough numbers, that by low redshift they have not grown sufficiently to alter the SFR from its value without AGN feedback.
In the fiducial simulation with $\rc = 0.1\,h^2m_\textrm{H}\,\textrm{cm}^{-3}$, however, the BHs form early enough, and in large enough numbers, that star formation is noticeably quenched also at low redshift, and brought in line with observations.

The influence of delaying the birth time of black holes is also clear in Fig. \ref{fig:rhombhz}, where on average, more massive black holes are produced with lower $\rc$.
This is because they have both more time to accrete gas, and more black holes with which to merge.
The trend is also seen in Fig. \ref{fig:rhombhsig}, which shows the $z=0$ $M_\textrm{BH}-\sigma$ relationship for these simulations.
We see that, in simulation F, more black holes have grown onto the observed relation by the present day, and display a smaller scatter than in simulations R1 or R2.

While $\rc$ influences the evolution of black holes and thus affects the cosmic SFR at high and low redshifts, its impact on the evolution of galaxies is minimal.
Figs. \ref{fig:rhoresig} and \ref{fig:rhobmstr} show, respectively, effective radius as a function of velocity dispersion and baryon fraction as a function of stellar mass for the simulated galaxies.
Neither indicates a strong trend with $\rc$, implying that the stellar and gas components of present-day galaxies are not influenced by when black holes form, at least by $z=0$.
This would seem to be in contradiction to Fig. \ref{fig:rhodensz}, which shows that the average gas density local to black holes depends on $\rc$, with higher values having higher gas densities (dotted lines).
However, this is simply due to the fact that black holes are only spawned in high density regions if $\rc$ is itself large; the trend is not indicative of the black holes' influence in those regions.

\section{Resolution Effects}
\label{sec:restests}

\begin{table*}
\caption{Input parameters and $z=0$ and $z=2$ quantities of the simulations for the study of resolution effects.
Measured quantities are given at $z=2$ for all simulations, and $z=0$ for those that were run to completion.
In all cases, values for the model parameters as in simulation F are assumed.
(1) Simulation name.
(2) Mass of dark matter particles.
(3) Initial mass of gas particles.
(4) Gravitational softening length.
(5) Stellar mass fraction $f_* = M_*/(M_*+M_\textrm{g}+M_\textrm{BH})$.
(6) Total number of black holes.
(7) Black hole mass fraction $f_\textrm{BH}=M_\textrm{BH}/(M_\textrm{BH}+M_\textrm{g}+M_\textrm{BH})$.
(8) Average ratio of black hole mass to stellar mass within $R_\textrm{e}$.
(9) Mass of most massive black hole.
(10) Redshift at which first black hole forms.}
\centering
\begin{tabular}[width=\textwidth]{cccccccccc}
	\hline
	Name & $M_\textrm{DM}$ & $M_\textrm{gas}$ & $\epsilon_\textrm{gas}$ & $f_*$ & $N_\textrm{BH}$ & $f_\textrm{BH}$ & $\left<M_\textrm{BH}/M_{*,R_\textrm{e}}\right>$ & $\max{(M_\textrm{BH})}$ & $z_i$\\
	& ($\times10^6\,h^{-1}\textrm{M}_\odot$) & ($\times10^6\,h^{-1}\textrm{M}_\odot$) & ($h^{-1}$ kpc) & & & ($\times10^{-5}$) & ($\times10^{-3}$) & ($\times10^7\textrm\,h^{-1}{M}_\odot$) &\\
	(1) & (2) & (3) & (4) & (5) & (6) & (7) & (8) & (9) & (10)\\
	\hline
	\multicolumn{10}{c}{$z=0$}\\
	\hline
	010mpc054 & 412 & 81.1 & 2.0 & 0.046 & 72 & 1.19 & 0.7 & 9.8 & 11.7\\
	010mpc096 & 73.4 & 14.4 & 1.125 & 0.053 & 206 & 1.36 & 0.6 & 7.4 & 16.0\\
	010mpc128 & 31.0 & 6.09 & 0.844 & 0.051 & 370 & 1.75 & 1.7 & 8.4 & 48.8\\
	\hline
	\multicolumn{10}{c}{$z=2$}\\
	\hline
	010mpc054  & 412 & 81.1 & 2.0 & 0.012 & 214 & 0.09 & 0.4 & 0.71 & 11.7\\
	010mpc096  & 73.4 & 14.4 & 1.125 & 0.020 & 643 & 0.53 & 1.1 & 0.89 & 16.0\\
	010mpc128 & 31.0 & 6.09 & 0.844 & 0.017 & 1,132 & 0.94 & 3.1 & 0.74 & 48.8\\
	010mpc256 & 3.87 & 0.76 &  0.422 & 0.017 & 7,517 &  6.43 & 30.0 & 2.83 & 48.8\\
	\hline
\end{tabular}
\label{tab:restest}
\end{table*}

\begin{figure}
	\includegraphics[width=0.47\textwidth,keepaspectratio]{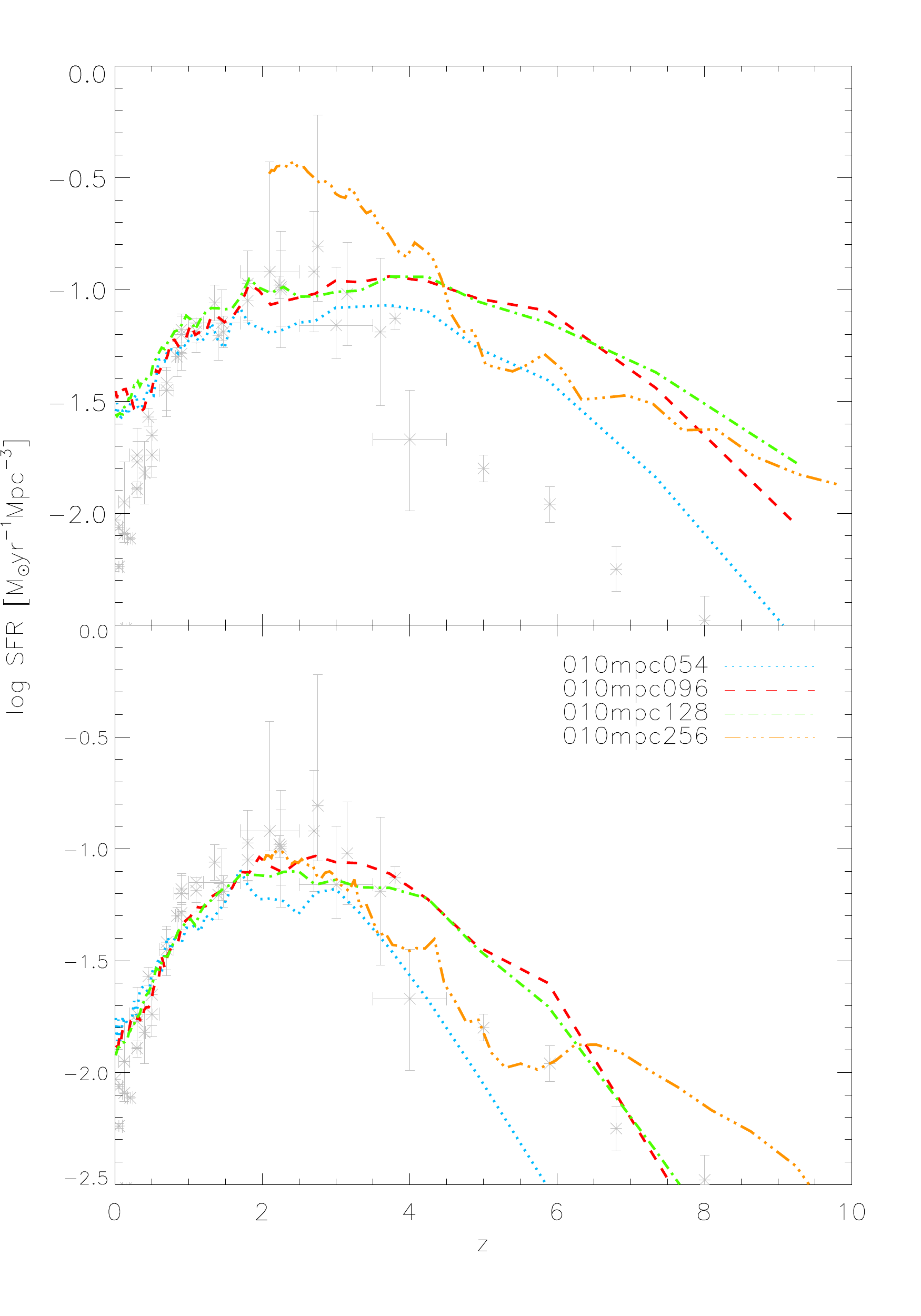}
	\caption{Cosmic star formation rate history for simulations of various resolutions without (top panel) and with (bottom panel) AGN feedback.
See Fig. 4 for the observational data sources.}
	\label{fig:sfrres}
\end{figure}

\begin{figure}
	\includegraphics[width=0.47\textwidth,keepaspectratio]{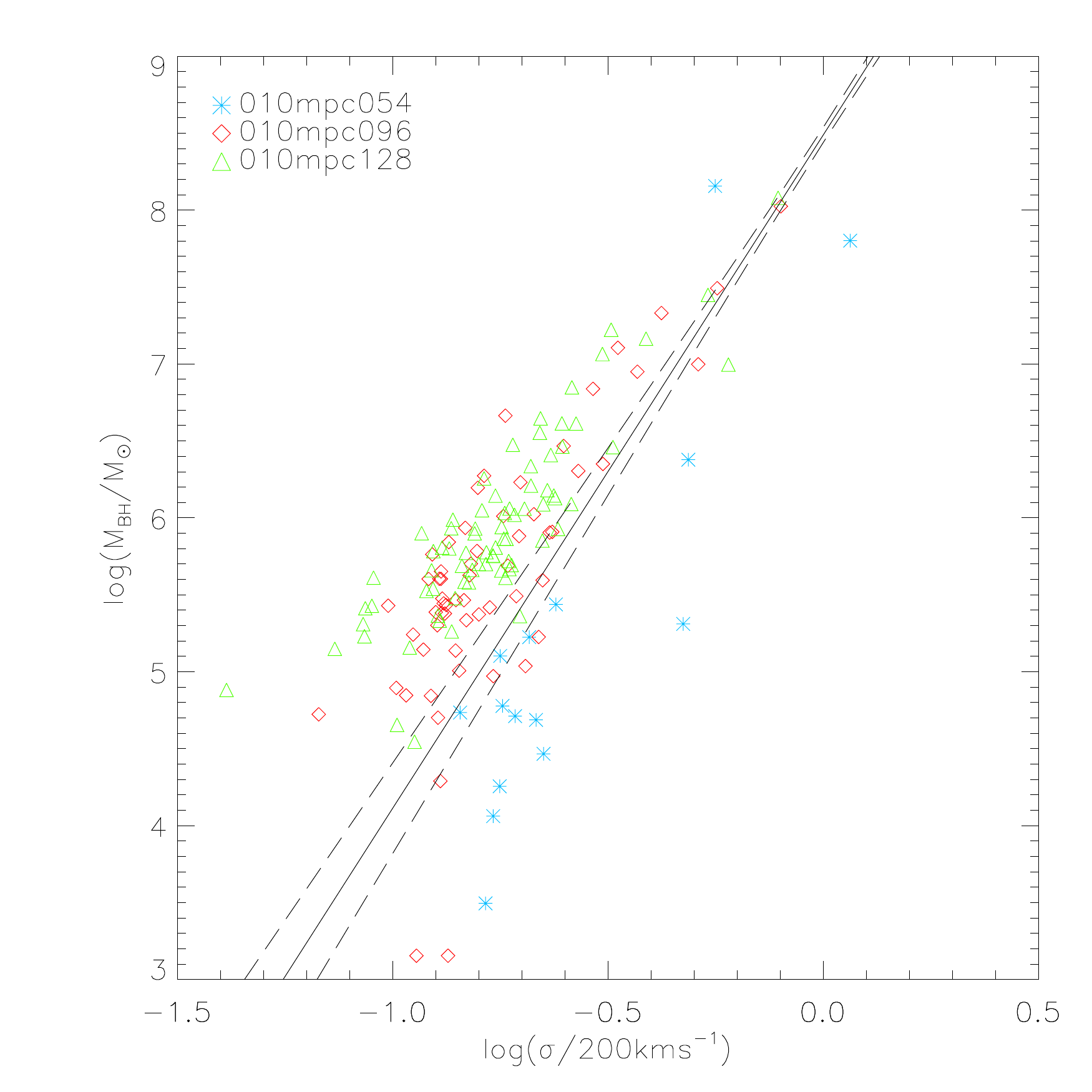}
	\caption{$M_\textrm{BH}-\sigma$ relation at $z=0$ for simulations of various resolutions.
	               The solid and dashed lines show the observed relation and scatter \citep{kormendyho13}, respectively.}
	\label{fig:msigres}
\end{figure}

\begin{figure}
	\includegraphics[width=0.47\textwidth,keepaspectratio]{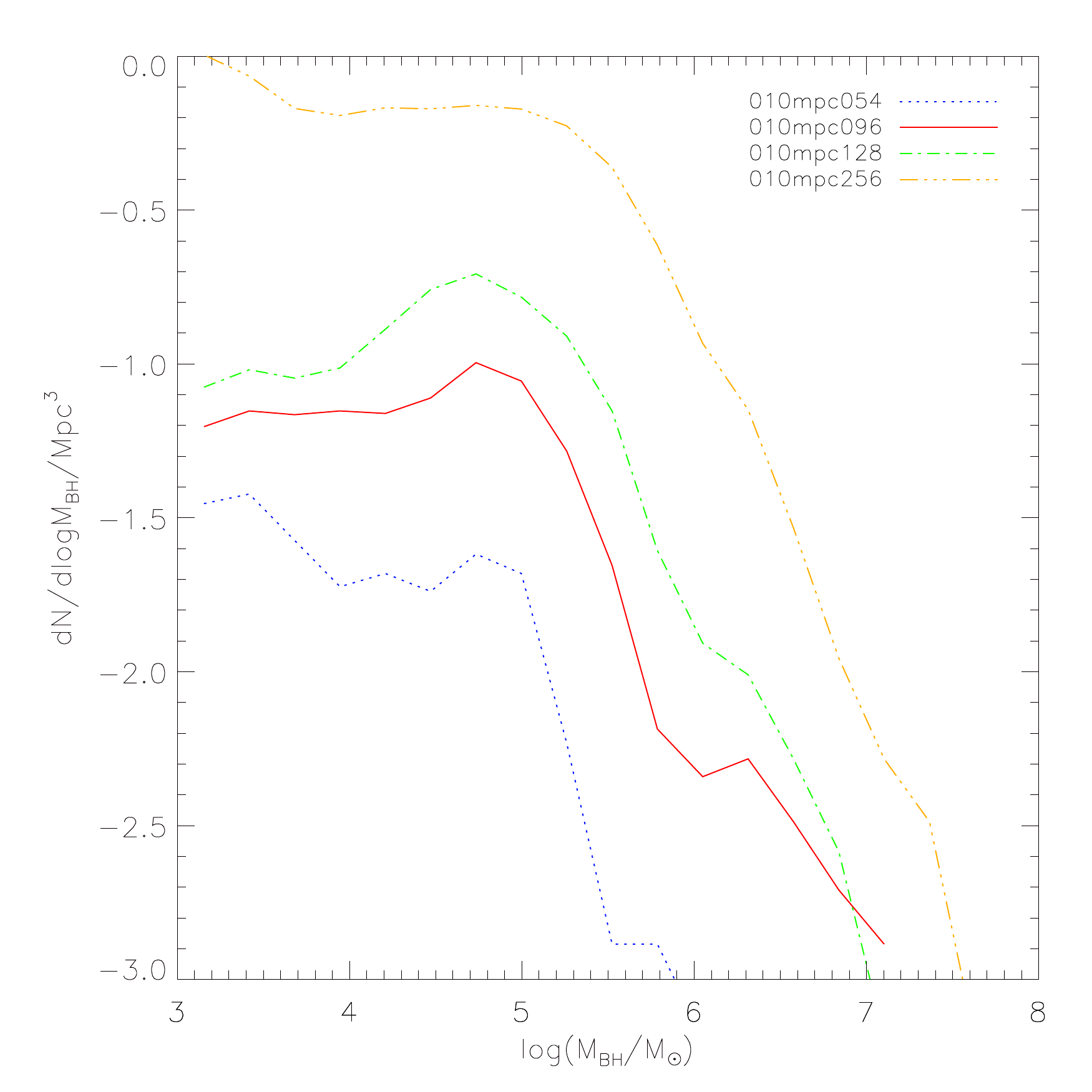}
	\caption{Black hole mass function at $z=2$ for simulations of various resolutions.}
	\label{fig:bhmfn}
\end{figure}

In this Section, we present the results of a group of simulations run at different resolutions in order to investigate the influence of resolution on our model.
In all cases, the parameters of the fiducial simulation of Section \ref{sec:ptests} are used, specifically $\alpha=1$, $\ef=0.25$, $\ms=10^3\,h^{-1}M_\odot$, and $\rc=0.1\,h^2m_\textrm{H}\,\textrm{cm}^{-3}$.
Simulations are given names of the form xxxmpcnnn, with xxx the comoving length of each side of the simulation box, and nnn the cube root of the initial number of both gas and dark matter particles (note that 010mpc096 is identical to simulation F of Section \ref{sec:ptests}).
The highest resolution simulation was run to $z=2$.
Therefore, we show in Table \ref{tab:restest} properties of all simulations at $z=2$, as well as the $z=0$ properties of those simulations that were run to completion.

As shown in Table \ref{tab:restest}, the stellar mass fractions are almost constant, and the cosmic SFRs at $z\lesssim2$  do not depend very much on the resolution (Fig. \ref{fig:sfrres}).
However, at high redshifts, some resolution dependencies are seen for both star and black hole formation.
With higher resolution, it is possible to resolve more high density regions at a given redshift, and therefore cosmic SFRs become higher at high redshift \citep[e.g.,][]{springel03,booth09}.
Fig. \ref{fig:sfrres} shows the cosmic SFRs in our simulations with various resolutions.
Without AGN feedback (top panel), higher resolution simulations show higher SFRs at $z\gtrsim2$.
For 010mpc256, supernova feedback also occurs at high redshift, which results in a fluctuating SFR history.
The SFR at high redshift could be reduced by adopting a high critical density for star formation \citep[e.g.,][]{guedes11}, although we do not use density as a star formation criterion in this model (Section \ref{sec:model}).
In our model, increasing $N_\textrm{FB}$ with resolution would mitigate against the fluctuations seen.
With our AGN model included (bottom panel), higher resolution simulations have more black holes at a given redshift, which provide more AGN feedback at high redshifts.
This is because more high density regions that satisfy equation \eqref{eq:bhform1} are resolved, and, with a given $N_\textrm{FB}$, more primordial gas particles remain, which satisfy equation \eqref{eq:bhform2}.
These resolution effects on star formation act in opposition, and in fact 010mpc96 and 010mpc128 show very similar cosmic SFRs (Fig. \ref{fig:sfrres}) with the same values of $\rc$ and $N_\textrm{FB}$.
However, some resolution dependence is still seen for our highest resolution simulation; 010mpc256 shows higher SFRs at $z>7$ and lower SFRs at $z=3-5$ than the lower resolution simulations.
In this case, convergence in the cosmic SFRs may be obtained with higher $\rc$ and/or $N_\textrm{FB}$.
Such a test is left to future works.


Fig. \ref{fig:msigres} shows the $z=0$ relation between $M_\textrm{BH}$ and $\sigma$, in which we see that there is good agreement across the simulations above $M_\textrm{BH}\sim10^6M_\odot$.
Below this mass, there is a slight offset, with 010mpc128 lying above the observed relation.
In higher resolution simulations, gas densities of a given value are resolved earlier, and are more abundant at a given redshift, so black holes are able to form earlier.
This then allows black holes more time to grow by gas accretion, as well as providing a larger population of black holes with which to merge.
Therefore, both the number and mass fraction of black holes are greater at $z=0$ in simulations with higher resolution, a trend mirrored at $z=2$.
We see in Fig. \ref{fig:bhmfn}, which shows the black hole mass function at $z=2$, that there is a trend for higher resolution simulations to contain a greater number of black holes.
This resolution dependence could be reduced by changing $\rc$ (see also Fig. \ref{fig:appbhmfn} in the Appendix).
A lower seed mass will also lower the $z=0$ mass of black holes so that they lie more closely on the observed $M_\textrm{BH}-\sigma$ relation (Fig. \ref{fig:msigres}).
$\alpha$ would not need changing, since the peak in SFR (Fig. \ref{fig:sfrres}) is already consistent with observations.
Therefore, for higher resolution, high $\rc$ or lower seed mass may be favoured for the $M_\textrm{BH}-\sigma$ relation.
We have not studied the effects of $N_\textrm{FB}$ on the $M_\textrm{BH}-\sigma$ relation.

\section{Discussion}
\label{sec:disc}
We have introduced a new AGN model, in which black holes are seeded at local peaks in gas density.
Because of this, the growth of black holes is correlated with hierarchical clustering of galaxies.
We have demonstrated that our model of the formation and evolution of black holes in cosmological simulations is able to reproduce a number of observations, including the $M_\textrm{BH}-\sigma$ relation and the cosmic star formation rate history.
The properties of galaxies, such as luminosity functions and colour magnitude relations, will be discussed in future papers.
Our model contains four parameters -- $\alpha$, $\ef$, $\ms$, and $\rc$ -- whose effects on the formation and evolution of black holes, stars, and gas have been investigated in Section \ref{sec:ptests}.
This has allowed us to determine a best set of parameters, in the sense of reproducing observations, with values $\alpha=1$, $\ef=0.25$, $\ms=10^3\,h^{-1}M_\odot$, and $\rc=0.1\,h^2m_\textrm{H}\,\textrm{cm}^{-3}$.

A seed mass of $10^3M_\odot$ is several orders of magnitude lower than the value typically used in other cosmological simulations involving the classical scheme of AGN feedback.
In principle, it might be possible to obtain a reasonably good fit to observations by simultaneously altering more than one of the model parameters from their fiducial values.
Based on the parameter dependencies found in Section \ref{sec:ptests}, we ran $\sim50$ parameter sets and did not find any better than the fiducial set.
We found that with a larger $\alpha$ (25--100) and smaller seed mass ($10^2M_\odot$), the cosmic SFR history and $M_\textrm{BH}-\sigma$ relation are still fairly well reproduced, but the sizes of galaxies tend to be even larger than observed.
The required seed mass of $\ms=10^{2-3}\,h^{-1}M_\odot$ suggests that the origin of the seed black holes is the deaths of Populations {\sc iii} stars (Section \ref{sec:model}).
Because no signatures of the chemical enrichment from PISNe have yet been observed (Section \ref{sec:intro}), we propose the following scenario: the masses of the first stars are $\lesssim 140 M_\odot$ due to fragmentation and suppressed accretion, which explode as core-collapse supernovae.
If the accretion is not suppressed, the mass of stellar cores grows beyond $\sim 300 M_\odot$, which become our seed black holes.

The value of $\ef=0.25$ is somewhat higher than has been necessary in earlier studies, which have used $\ef$ in the range 0.05 \citep{springel05,sijacki07,dimatteo08} to 0.15 \citep{booth09} in order to reproduce the observed $M_\textrm{BH}-\sigma$ relation, but see also \citet{figueroa13} who adopt $\ef=0.2$ or $0.8$, depending on the accretion rate.
Additionally, observations of massive molecular outflows from the quasar Mrk 231 suggest $0.02\lesssim \ef \lesssim 0.06$ \citep{feruglio10,cicone12}.
In our simulations, a constant $\ef$ is applied for all gas at all times.
Ultimately, the value of $0.25$ found here should be viewed as a rough average used in lieu of a more accurate expression for $\ef$ based on the properties of the gas around black holes.
In reality, the extent to which radiation couples to gas will depend on various factors, including the density and metallicity of gas.
In addition, we only allow the radiation to affect the thermal energy of the gas; there is no kinetic component, nor do we employ the radiative electromagnetic feedback of \citet{vogelsberger13}.
If the gas is more easily removed from galaxy centres by kinetic coupling, this can reduce the value of $\ef$.
However, \citet{dimatteo08} argued that the mode of coupling is largely irrelevant as long as the feedback energy is distributed in a volume small compared to that of the galaxy, and in a time shorter than the dynamical time of the galaxy.
Indeed, even without kinetic feedback, outflows are seen from massive galaxies in our simulations.
In the most massive galaxies, we find that the wind speed is $\gtrsim100$kms$^{-1}$ at the edge of galaxies, and can be higher near the center as observed \citep[e.g.][]{sturm11,cicone12,maiolino12,veilleux13}.

\begin{figure}
	\includegraphics[width=0.47\textwidth,keepaspectratio]{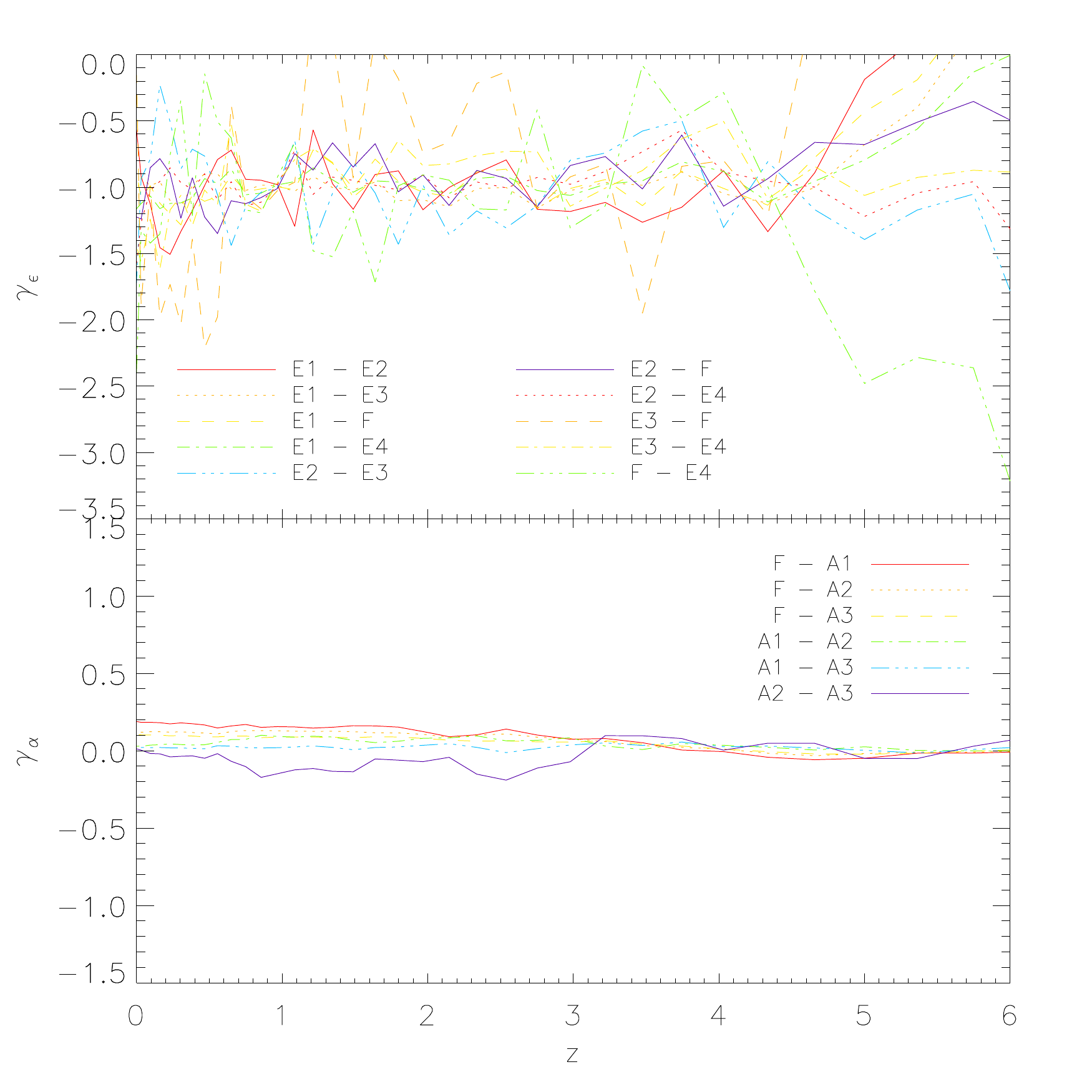}
	\caption{\emph{Top panel}: Evolution of $\gamma_\epsilon$ (see equations \eqref{eq:gam1} and \eqref{eq:gam2}) with redshift for all pairings of the simulations F and E1-3.
		    \emph{Bottom panel}: Evolution of  $\gamma_\alpha$ (see equations \eqref{eq:gam3} and \eqref{eq:gam4}) with redshift for all pairings of the simulations F and A1-3.}
	\label{fig:gamma}
\end{figure}

That the value of $\ef$ has no bearing on the cosmic star formation history is surprising, as star formation is directly related to the internal energy of gas particles through the rapid cooling condition.
However, this can be explained by the self-regulation of the black hole growth as follows: the simulations F and E1-4 are identical, apart from their value of $\ef$, which means that, to a good approximation, black holes form at the same locations, at the same times, and with the same masses in each.
Assuming a power law dependence of $\dot M_\textrm{BH}$ on $\ef$, we may write
\begin{equation}\label{eq:gam1}
	\dot M_\textrm{BH} = k\ef^{\gamma_\epsilon},
\end{equation}
and calculate $\gamma_\epsilon$ as
\begin{equation}\label{eq:gam2}
	\gamma_\epsilon = \frac{\log(\overline{\dot M}_\textrm{BH,1} / \overline{\dot M}_\textrm{BH,2})}{\log(\epsilon_\textrm{f,1} / \epsilon_\textrm{f,2})},
\end{equation}
where $\overline{\dot M}_\textrm{BH,i}$ denotes the mean black hole accretion rate of the simulation with $\ef = \epsilon_\textrm{f,i}$.
We show $\gamma_\epsilon$ as a function of redshift for pairs of $\ef$ in the top panel of Fig. \ref{fig:gamma}.
Although there is some scatter, we see that $\gamma_\epsilon \sim -1$ for $z<4$ (when gas accretion is the dominant growth mode) in all cases.
From equation \eqref{eq:efb}, we see that $\dot M_\textrm{acc} \propto \ef^{-1}$ implies that $E_\textrm{FB}$ is independent of $\ef$, and so that black hole growth is self-regulated.

The self-regulated growth also explains the independence of black hole mass on the value of $\alpha$.
Again, when and where black holes form is unaffected by the value of $\alpha$.
As before, we assume a power law dependence of $M_\textrm{BH}$ on $\alpha$:
\begin{equation}\label{eq:gam3}
	M_\textrm{BH}=k\alpha^{\gamma_\alpha},
\end{equation}
and calculate $\gamma_\alpha$ as
\begin{equation}\label{eq:gam4}
	\gamma_\alpha = \frac{\log(\overline{M}_\textrm{BH,1}/\overline{M}_\textrm{BH,2})}{\log(\alpha_\textrm{1}/\alpha_\textrm{2})},
\end{equation}
where $\overline{M}_\textrm{BH,i}$ denotes the mean black hole mass of the simulation with $\alpha=\alpha_\textrm{i}$.
We show $\gamma_\alpha$ as a function of redshift for pairs of $\alpha$ in the bottom panel of Fig. \ref{fig:gamma}.
It is clear that $\alpha\sim0$ at all times, and so black hole mass is unaffected by $\alpha$.
From equation \eqref{eq:efb}, we see that the total energy, $E$, a black hole radiates to gas particles is approximately given by
\begin{equation}
	E=\er \ef M_\textrm{acc}c^2,
\end{equation}
$M_\textrm{acc}$ being the total accreted gas mass.
Since $M_\textrm{BH}$ does not depend on the value of $\alpha$, we can see that $E$ is constant, and hence that the growth is self-regulated.

Our adopted `best' value for the minimum gas density for black holes to form, $\rc=0.1\,h^2m_\textrm{H}\,\textrm{cm}^{-3}$, is not intended as a physical density threshold of the gas that would have formed the first black holes in the universe, because the resolution of our cosmological simulations is insufficient to resolve such small scale physics.
Rather, it sets the timescale, for a given resolution, at which black holes start to form as large scale structure collapses.
In higher resolution simulations, a lower value of $\rc$ causes black holes to form earlier and in greater numbers, as shown in Section \ref{sec:restests}.
In the simulations presented in Section \ref{sec:restests}, we apply the same $N_\textrm{FB}$ for both supernova and AGN feedback as in Section \ref{sec:ptests}, which ensures the same total mass in the feedback region at a given resolution, but results in a smaller region for higher resolution simulations.
By adopting a resolution dependent value for $N_\textrm{FB}$ such that metals are distributed over the same distance, the number of black holes will be reduced.

A major cause of uncertainty in our model is our prescription for calculating the accretion rate of gas onto black holes.
We use the Bondi-Hoyle accretion rate, as in many previous works \citep[e.g.][]{springel05,dimatteo08,booth09}.	
This formulation is for spherical accretion in the Newtonian regime.
The angular momentum of infalling gas is ignored, which could, more realistically, cause the gas to settle into a disk around the black hole, rather than being accreted by it.
Jets may form with a strong magnetic field \citep{tchekhovskoy11}, which may help remove angular momentum.
Additionally, the gravitational potential felt by the gas is assumed to be dominated by the black hole, whereas that of the galaxy will be more important in these simulations.
Attempts to address some of these issues have been made; for example, in \citet{power11} a two-component `accretion disk particle' is used that models both a black hole and its accretion disk.
Gas is accreted onto the disk if it comes within some radius $R_\textrm{acc}$ that is of order the gravitational softening length.
\citet{angles13}, on the other hand, calculate an accretion rate based on angular momentum transport via gravitational torques within the central $\sim 1\,$ kpc of galaxies.
Such methods require a somewhat higher spatial resolution than in this paper, and we will leave these to a future study.

\section{Conclusions}
\label{sec:conc}

We have presented a new model for the formation of black holes in cosmological simulations, motivated by the first star formation.
Black holes are formed from metal free gas of high enough density.
They may then grow via both mergers and through gas accretion.
In this model, black holes are formed anywhere in the simulation volume with gas satisfying the above criteria.
We therefore do not restrict black holes to form only within dark matter haloes, as some previous models have, but still find that most galaxies contain at least one black hole by the present.
Massive black holes heat the surrounding material, suppressing star formation at the centres of galaxies, and driving galactic winds.

We constrain the parameters of the model by comparing the simulations to observations, and successfully reproduce the black hole mass -- galaxy mass relation, the cosmic star formation rate history, and the mass -- size relation of galaxies.
The best model has $\alpha=1$, $\ef=0.25$, $\ms=10^3\,h^{-1}M_\odot$, and $\rc=0.1\,h^2m_\textrm{H}\,\textrm{cm}^{-3}$ (Section \ref{sec:ptests}).
For high resolution, however, a higher value of critical density for black hole formation, $\rc$, a larger number of feedback neighbours, $N_\textrm{FB}$, and/or slightly smaller seed black holes may be favoured (Section \ref{sec:restests}).

The mass with which black holes are seeded is $\sim10^3\,h^{-1}M_\odot$ in order to match observations fully, though there are sets of parameters with $\ms=10^2\,h^{-1}M_\odot$ that can reproduce observations fairly well.
The required seed mass of $\ms=10^{2-3}\,h^{-1}M_\odot$ suggests that the origin of the seed black holes is the deaths of Population {\sc iii} stars; the masses of the first stars are $\gtsim 140M_\odot$ due to fragmentation and suppressed accretion, which explode as core-collapse supernovae.
If the accretion is not suppressed, the mass of stellar cores grows beyond $300M_\odot$, which become our seed black holes.

We found that black hole mass did not depend on the accretion rate modifier, $\alpha$, and that SFR did not depend on the fraction of radiation that coupled to gas, $\ef$.
This suggests that black hole growth is self-regulated; the same amount of energy is radiated, or equivalently, the same mass of gas is accreted by black holes regardless of the values of these parameters (Section \ref{sec:disc}).

\section*{Acknowledgements}
We would like to thank M. Hardcastle, T. Di Matteo, G. Bicknell, J. Silk, and A. Hopkins for useful discussions.
PT acknowledges funding from an STFC studentship, and thanks S. Lindsay for fruitful discussions.
This work has made use of the University of Hertfordshire Science and Technology Research Institute high-performance computing facility.
Finally, we thank V. Springel for providing Gadget-3.


\footnotesize{
\bibliographystyle{mn2e}
\bibliography{./refs}
}


\normalsize
\appendix
\section{Black Hole Mass Function \& Mass Density}
\label{sec:appen}

\begin{figure*}
	\centering
	\includegraphics[width=0.9\textwidth,keepaspectratio]{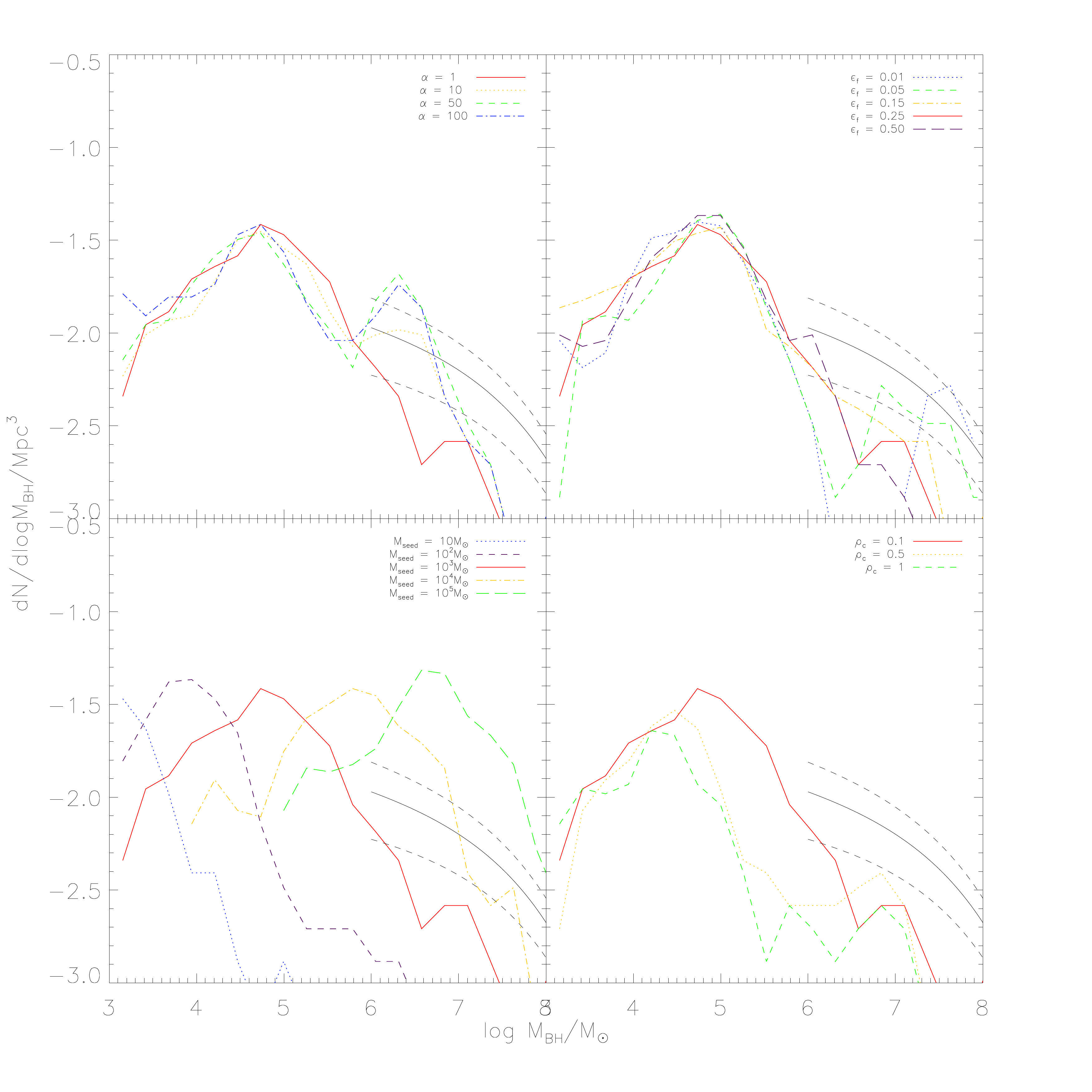}
	\caption{Black hole mass functions for the simulations presented in Section \ref{sec:ptests}.
		    In all panels, the observed trend and scatter from \citet{shankar04} are plotted as the solid and dashed black lines, respectively, for $\log M_\textrm{BH}/M_\odot \ge 6$, the range in which their fit holds, and the fiducial model is shown as the solid red line.
		    From top to bottom and left to right, model parameters are held at their fiducial values except for $\alpha$, $\ef$, $\ms$, and $\rc$ respectively.}
	\label{fig:appbhmfn}
\end{figure*}

\begin{figure*}
	\centering
	\includegraphics[width=0.9\textwidth,keepaspectratio]{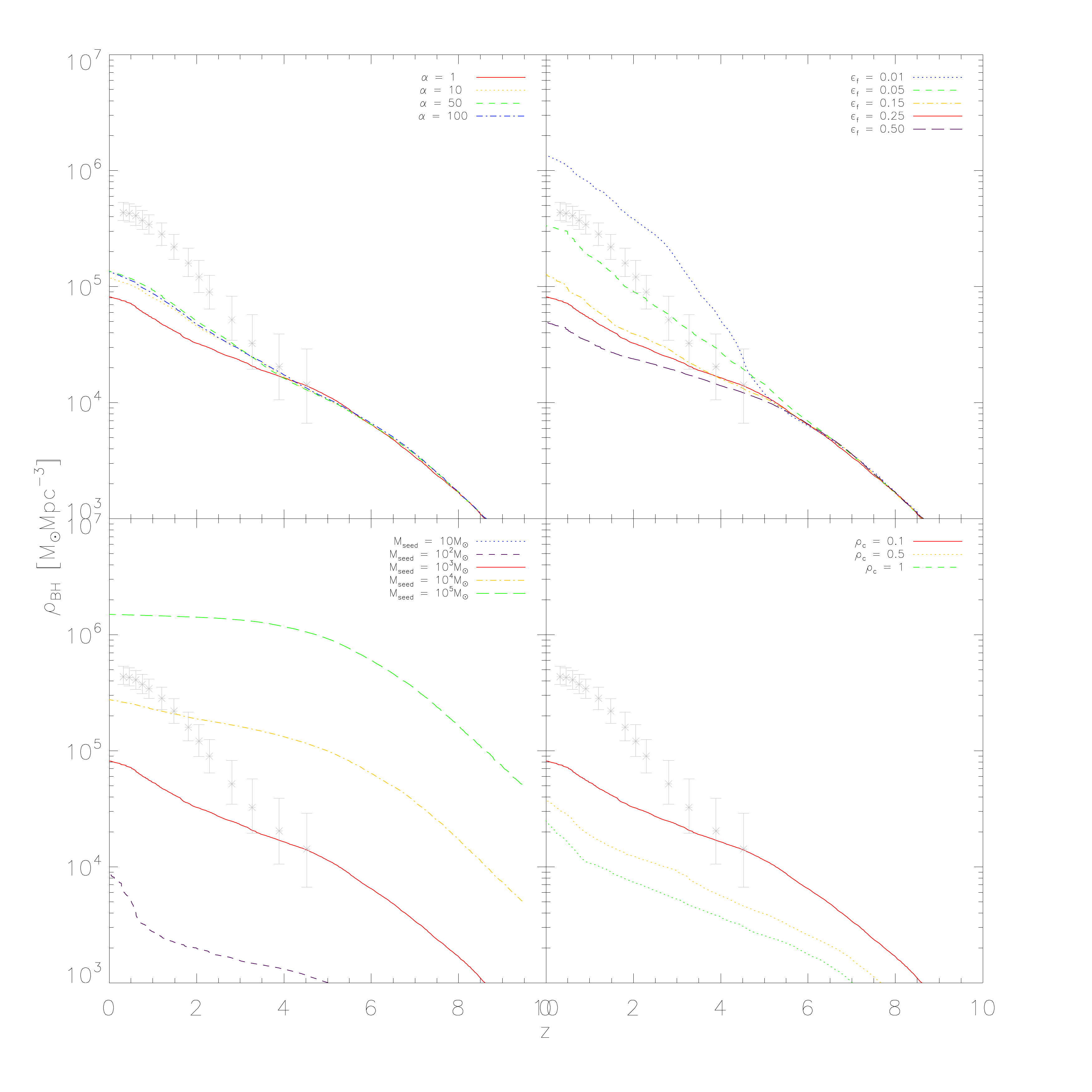}
	\caption{Black hole mass density as a function of redshift for the simulations presented in Section \ref{sec:ptests}.
		    In all panels, the observational data are from \citet{hopkins07}, and the fiducial model is shown as the solid red line.
		    From top to bottom and left to right, model parameters are held at their fiducial values except for $\alpha$, $\ef$, $\ms$, and $\rc$ respectively.}
	\label{fig:appbhrho}
\end{figure*}

Although the volume of our simulation box is not large enough to discuss these, we show the $z=0$ black hole mass function, and the redshift evolution of black hole mass density, for the different simulations presented in Section \ref{sec:ptests}, in Figs. \ref{fig:appbhmfn} and \ref{fig:appbhrho} respectively.
Observations of the mass function are constrained only at $\log M_\textrm{BH}/M_\odot \ge 6$ \citep{shankar04}.
From the top left panel of Fig. \ref{fig:appbhmfn}, there is a slight trend for more massive black holes to form in simulations with larger $\alpha$, while the top right panel shows that the value of $\ef$ does not influence the mass of the black holes, as found in Section \ref{sec:ptests}.
Unsurprisingly, the bottom left panel demonstrates that a more massive seed mass produces more massive black holes by $z=0$, and $\ms=10^{4-5}$ seems to give too large a number of black holes at $10^{6-7}M_\odot$ with our AGN model.
In the bottom right panel, we see that lower $\rc$ produces a larger number of black holes of mass $10^{5-6}M_\odot$, and at $10^6M_\odot$, our best parameter set (solid line) gives a number of black holes as large as observed.

We see in Fig. \ref{fig:appbhrho} the same dependence of black hole mass on the model parameters as before.
As in Section \ref{sec:ptests}, $\ms=10^5\,h^{-1}M_\odot$ sems too large to be consistent with observations, and at $z \sim 3-5$, our best parameter set (solid line) gives a black hole density as large as observed.
Since the most massive black holes are not present in our simulations, the black hole mass density tends to be lower than observed.

\bsp

\label{lastpage}

\end{document}